# Analytical modeling of coated plasmonic particles

Nikolai G. Khlebtsov[1,2,*], Sergey V. Zarkov[1,3]

[1]Institute of Biochemistry and Physiology of Plants and Microorganisms, "Saratov Scientific Centre of the Russian Academy of Sciences," 13 Prospekt Entuziastov, Saratov 410049, Russia

[2]SaratovStateUniversity, 83 Ulitsa Astrakhanskaya, Saratov 410012, Russia

[3]Institute of Precision Mechanics and Control, "Saratov Scientific Centre of the Russian Academy of Sciences," 24 Ulitsa Rabochaya, Saratov 410028, Russia

[*]To whom correspondence should be addressed. E-mail: (NGK) khlebtsov@ibppm.ru

**Abstract.** Biomedical applications of plasmonic nanoparticle conjugates need control over their optical properties modulated by surface coating with stabilizing or targeting molecules often attached to or embedded in the secondary functionalization shell, such as silica. Although current numerical techniques can simulate the plasmonic response of such structures, it is desirable in practice to have analytical models based on simple physical ideas that can be implemented without considerable computer resources. Here, we present two efficient analytical methods based on improved electrostatic approximation (IEA) and modal expansion method (MEM) combined with the dipole equivalence method (DEM). The last approach avoids additional electromagnetic simulations and provides a direct bridge between analytical IEA and MEM models for bare particles and those with multilayer shells. As simple as the original IEA and MEM, the developed analytical extensions provide accurate extinction and scattering spectra for coated particles compared to exact calculations by separation of variable method and COMSOL. The possibility and accuracy of analytical models are illustrated by extensive simulations for prolate and oblate gold and silver nanoparticles with a maximal size of up to 200 nm, aspect ratio from 2 to 6, and 3-30 nm dielectric coating.

## 1. Introduction

The use of plasmonic nanoparticles (NPs) in biomedicine is based on their functionalization with various stabilizing ligands and biomolecules that provide various functional properties.[1,2] For example, coating the surface of gold nanoparticles (AuNPs) with thiolated PEG molecules ensures their stabilization in the bloodstream and the protection of immune system cells.[3] Coating NPs with antibodies or aptamers ensures the specific interaction of conjugates with biotarget sites.[4] In addition, already during the synthesis process, NPs are often coated with molecules of a particular stabilizer presented in the reaction mixture. A typical example is the hexadecyltrimethylammonium bromide (CTAB) layer on the surface of gold nanorods.[5] Thus, from a structural point of view, NP conjugates are always at least bilayer particles with a plasmonic core and a dielectric shell.

Modern numerical methods can predict the optical properties of bilayer (or multilayer) particles.[6] However, for many practical purposes, it is desirable to have simple analytical approximations that can be easily applied to realistic models that consider the random orientation and polydispersity of particles in size and shape. The simplest approximation is based on the small particle size compared to the light wavelength and uses electrostatic approximation (EA).[7] However, this approximation has a limited range of applicability regarding the size of plasmonic nanoparticles, especially silver ones. Le Ru *et al*. developed a T-matrix expansion in powers of the diffraction size of spheroidal particles,[8] which was used to formulate the improved electrostatic approximation (IEA).[9] While analytically straightforward as EA, the improved IEA approximation has a broader range of applicability regarding particle parameters. However, this method has not yet been applied to single- or multilayer-coated plasmonic nanoparticles.

To go beyond spheroidal shape limitations, García de Abajo and co-workers[10] developed universal analytical modeling for plasmonic nanoparticles by using the modal expansion method (MEM) based on electrostatic eigenmode expansion accomplished by size and shape-dependent correction for the retardation and radiation damping effects. With benchmark numerical solutions, they showed high accuracy of MEM for a broad shape family of gold and silver NPs.

In this work, for the first time to our knowledge, we describe an extension of the IEA and MEM methods to plasmonic NPs with a dielectric coating. The capabilities of the new version of IEA are illustrated by calculating the extinction and scattering spectra of bilayer gold and silver nanoparticles with random orientations. Models of prolate and oblate spheroids describe the particles' shape, and the accuracy of the of the new IEA version is assessed by comparison with calculations using the separation of variables method (SVM)[11] and the commercial COMSOL package. The capabilities of the new MEM version are illustrated by the example of gold nanorods coated with a dielectric layer. We show excellent accuracy of MEM in comparison with numerical COMSOL calculations.

## 2. Methods

Electrodynamic full-wave finite-element simulations were carried out with the commercial package COMSOL Multiphysics 5.1 (Wave Optics module). These simulations were made using a standard 3D option and a new version called 2.5D formalism[12] that is applicable to axially symmetrical scatterers. In 2.5D formalism, the incident and scattered fields are expanded in cylindrical harmonics, and the response of each harmonic is computed separately, resulting in considerable computational savings in terms of memory and processing time.[13] Compared to 3D simulations, the 2.5D formalism allows for averaging cross-sections over random particle orientations with acceptable computation burdens.

We used SVM codes[11] (C-M_SPH version 3.3i, kindly provided by N. V. Voshchinnikov) for confocal two-layered spheroids. SVM calculates the averaged cross-sections faster than 2.5D COMSOL, but its accuracy was found to be limited in the case of oblate particles. Fastest codes for randomly oriented ensembles were obtained with the T-matrix method described previously for a general case of absorbing hosts.[14] However, those T-matrix codes are applicable to homogeneous particles only, thus making it impossible to compare cross-sections for coated particles. Therefore, we used 3D and 2.5D COMSOL calculations as benchmarks for this work.

## 3. Improved electrostatic approximation of coated spheroids

### 3.1. Model

We consider a two-layer spheroid in which the semi-axes and permittivities of the core and shell are equal to $(a_1 = b_1, c_1)$, $\varepsilon_1$ and $(a_2 = b_2, c_2)$, $\varepsilon_2$, respectively. A homogeneous and isotropic environment is characterized by a dielectric function $\varepsilon_m$, which can be complex for an absorbing host. Due to the condition of confocality of the core and shell, the semi-axes of spheroids in the standard surface equation $(x^2 + y^2)/a_i^2 + z^2/c_i^2 = 1$ obey the relations

$$a_2^2 = a_1^2 + s^2, \quad c_2^2 = c_1^2 + s^2. \tag{1}$$

A similar model was considered when solving the problem in the EA approximation[7] and by SVM.[11] It should be emphasized that the parameter $s$ is not equal to a constant thickness of the shell since the thicknesses of the shell along $c_2 - c_1$ and across $a_2 - a_1$ the symmetry axis are different. To describe a more or less uniform distribution of shell material, we define the thickness parameter as the geometric mean of the thicknesses $c_2 - c_1$ and $a_2 - a_1$

$$s = \sqrt{(c_2 - c_1)(a_2 - a_1)}. \tag{2}$$

Thus, the geometric model of a two-layer particle is determined by three parameters: the semi-axes of the core $a_1, c_1$ and the shell thickness parameter $s$ according to equation (2). For simplicity, we will call the parameter $s$ the shell thickness in what follows. By solving equations (1) and (2), the semi-axes of the shell $a_2, c_2$ can be found for given $a_1, c_1$ and $s$.

The dielectric constants of gold were taken from the work.[15] For silver, the spline described in the SI was used (Section S3). One has to emphasize here that we use the bulk optical constants to predict the optical response of small nanoparticles whose material constants can differ from their bulk counterparts due to size-limiting effects. In terms of the Drude model, the bulk damping constant of free electrons should be modified by the addition of three possible contributions related to radiation damping,[16] surface electron scattering,[17,18] and chemical interface damping.[19-21] In the IEA, radiation damping is always included in electromagnetic response by definition, thus leaving only two other damping mechanisms subject to consideration. All the above damping mechanisms lead to the well-known decrease in plasmonic peaks and their broadening. There are no problems in the dimensional correction of the Drude model by properly modifying the effective path length of electrons. However, our goal here is to examine IEM and MEM models compared to exact benchmark simulations. Therefore, we restricted ourselves by considering the dielectric functions of bulk metals.

Related to the above point, the dielectric coating of plasmonic particles does not have to be uniform.[22,23] For example, CTAB has a chemical interaction at the Au nanorod surface and an alkane tail. Thus, our model with a constant dielectric function of nanoparticle coating should be considered a first approximation. More accurate treatment should consider a combined chemical interaction at the Au surface and the dielectric function of the alkane layer, including possible interactions at the ammonium-metal interface. We have demonstrated previously how the non-uniform biopolymer distribution near gold nanosphere conjugates modulates their optical properties.[24] However, such subtle modifications appear redundant, at least at this stage of analytical model development.

The refractive index of the shell $n_2 = \sqrt{\varepsilon_2} = 1.5$ was chosen to be close to that of many biopolymers, including typical stabilizers such as CTAB or hexadecyltrimethylammonium chloride (CTAC). In all calculations, the refractive index of water was used as the external medium (see SI, Section S3).

### *3.2. Improved electrostatic polarizability of coated spheroids*

The following relations determine the polarizability of a homogeneous spheroid in IEA[9]

$$\alpha_{1i} = \beta_{1i} F_{1i}(\varepsilon_1, \varepsilon_m), \quad i = a, c\ (x, z), \tag{3}$$

$$F_{1i}(\varepsilon_1,\varepsilon_m)=\frac{1}{1-\Omega_{1i}X_1^2-i\dfrac{2}{3h_1^2}\beta_{1i}X_1^3},\quad i=a,c\,(x,z), \tag{4}$$

where

$$\beta_{1i}\equiv\alpha_{1i}^{EA}/R_1^3=\frac{\varepsilon_1-\varepsilon_m}{3\varepsilon_m+3L_{1i}(\varepsilon_1-\varepsilon_m)} \tag{5}$$

is the normalized electrostatic polarizability of a homogeneous spheroid, $V_1=(4\pi/3)a_1^2c_1\equiv(4\pi/3)R_1^3$ and $h_1=c_1/a_1$ are the volume and aspect ratio of the metallic core, respectively.

In formulas (3)-(5), it is assumed that the symmetry axis is directed along the axis of the coordinate system associated with the particle, the first index 1 corresponds to the metallic core, and the second index $i=a,c\ (x,z)$ corresponds to the direction of the electric field across or along the symmetry axis. The size of the semi-major axis of the metal core determines the size parameter $X_1=k_mc_1=(2\pi/\lambda)n_mc_1$. Formulas for geometric depolarization factors $L_{1i}$ and functions $\Omega_{1i}$ are given in Refs.[8,9] and are provided in the SI for convenience (Section S1).

To determine the polarizability of a coated spheroid, we use the dipole equivalence method (DEM).[25] The only difference from the previously described approach[25] is that electrostatic polarizability (5) is not used for the metal core, but IEA polarizability is used according to (3) and (4). The reason for this choice is significantly higher accuracy and a wider range of applicability for particle parameters when calculating the integral absorption and scattering cross sections.[8,9] In this modification, the method for calculating the IEA polarizability of a coated spheroid consists of four steps (Figure 1).

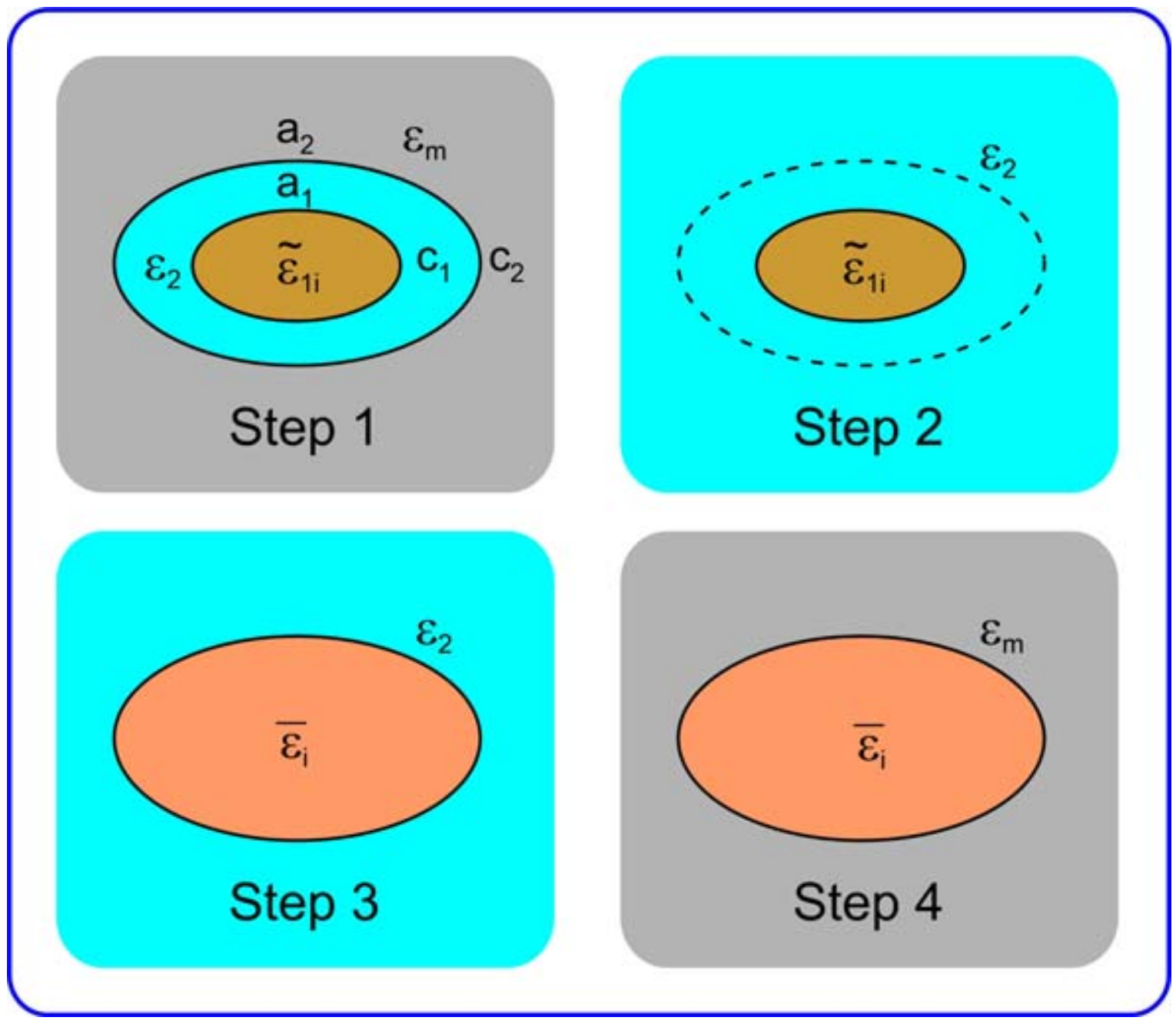

**Figure 1**. Scheme for calculating the IEA polarizability of coated plasmonic nanoparticles using DEM.

In the first step, we replace the IEA polarizability (3) with the equivalent EA polarizability of the spheroid (5) with the dielectric tensor $\tilde{\varepsilon}_{1i}$:

$$\frac{\tilde{\varepsilon}_{1i}-\varepsilon_m}{\varepsilon_m+L_{1i}(\tilde{\varepsilon}_{1i}-\varepsilon_m)}=F_{1i}(\varepsilon_1,\varepsilon_m)\frac{\varepsilon_1-\varepsilon_m}{\varepsilon_m+L_{1i}(\varepsilon_1-\varepsilon_m)}\equiv B_{1i}\,. \tag{6}$$

By solving Eq. (6), we find the principal values of the tensor

$$\tilde{\varepsilon}_{1i}=\varepsilon_m\left(1+\frac{B_{1i}}{1-L_{1i}B_{1i}}\right). \tag{7}$$

Next, we use DEM to determine the principal values of the dielectric polarizability tensor for coated particles. Specifically, the particle with the dielectric function (7) in the second step is placed in an isotropic medium with a dielectric constant of the coating $\varepsilon_2$. If we surround the initial particle with an imaginary confocal spheroid, $(a_2=b_2,c_2)$, then the particle's polarizability will not change. In the third step, we replace the imaginary particle in the medium $\varepsilon_2$ (step 2) with an equivalent particle with an average dielectric constant $\bar{\varepsilon}_i$ (step 3). Equating the polarizabilities of particles at steps 2 and 3, we obtain an equation for determining the principal values of the average permittivity tensor $\bar{\varepsilon}_i$ of the averaged auxiliary particle

$$R_1^3 \frac{\tilde{\varepsilon}_{1i} - \varepsilon_2}{\varepsilon_2 + L_{1i}(\tilde{\varepsilon}_{1i} - \varepsilon_2)} = R_2^3 \frac{\overline{\varepsilon}_i - \varepsilon_2}{\varepsilon_2 + L_{2i}(\overline{\varepsilon}_i - \varepsilon_2)}, \qquad (8)$$

where $R_2$ is the equivalent volume radius, $V_2 = (4\pi/3)a_2^2 c_2 = (4\pi/3)R_2^3$; $L_{2i}$ are the depolarization factors of a spheroid with the shell semi-axes. By analogy with solution (7), from equation (8), we find the principal values of the tensor $\overline{\varepsilon}_i$ across and along the symmetry axis ($i = a, c\ (x, z)$):

$$\overline{\varepsilon}_i = \varepsilon_2 \left( 1 + \frac{A_{1i}}{1 - L_{2i} A_{1i}} \right), \qquad (9)$$

$$A_{1i} = f \frac{\tilde{\varepsilon}_{1i} - \varepsilon_2}{\varepsilon_2 + L_{1i}(\tilde{\varepsilon}_{1i} - \varepsilon_2)}, \qquad (10)$$

where $f = (R_1 / R_2)^3$ is the core volume fraction for the coated particle. In the last fourth step, the auxiliary particle with permeability (9) is placed in the host medium $\varepsilon_m$, which gives the averaged IEA polarizability of a coated particle placed in the host:

$$\overline{\alpha}_i^{IEA} = R_2^3 \frac{\overline{\varepsilon}_i - \varepsilon_m}{3\varepsilon_m + 3L_{2i}(\overline{\varepsilon}_i - \varepsilon_m)}. \qquad (11)$$

Polarizability (11), obtained in the fourth step, is equivalent to the polarizability of the original coated spheroid in the host medium. The described scheme can be applied to a particle with an arbitrary number of different coatings (see below section 4.3). One important note is in order here. Suppose we omit step 1 and use the metal dielectric function $\varepsilon_1$ instead of $\tilde{\varepsilon}_{1i}$. Then, the solution (11) will be completely equivalent to the electrostatic polarizability of a coated spheroid obtained from the solution of Laplace's equation.[7]

A simpler scheme is possible in which the DEM is applied directly to a particle with IEA polarizability $\alpha_{1i} = \beta_{1i} F_{1i}$. In the first step, we place a particle with such polarizability in a dielectric-coated medium $\varepsilon_2$ and surround the particle by an imaginary confocal spheroid with shell semi-axes $(a_2 = b_2, c_2)$. Next, in the second step, we replace the imaginary particle in the medium with an equivalent particle with an average permittivity tensor $\overline{\varepsilon}_i$. Now, for $\overline{\varepsilon}_i$ permittivity, instead of equation (8), we get

$$R_1^3 F_{1i}(\varepsilon_1,\varepsilon_2)\frac{\varepsilon_1-\varepsilon_2}{\varepsilon_2+L_{1i}(\varepsilon_1-\varepsilon_2)} = R_2^3\frac{\overline{\varepsilon}_i-\varepsilon_2}{\varepsilon_2+L_{2i}(\overline{\varepsilon}_i-\varepsilon_2)}, \tag{12}$$

and further

$$\overline{\varepsilon}_i = \varepsilon_2\left(1+\frac{D_i}{1-L_{2i}D_i}\right), \tag{13}$$

$$D_i = fF_{1i}(\varepsilon_1,\varepsilon_2)\frac{\varepsilon_1-\varepsilon_2}{\varepsilon_2+L_{1i}(\varepsilon_1-\varepsilon_2)}. \tag{14}$$

In the third step, the auxiliary particle with permittivity (13) is placed in the host medium with permittivity $\varepsilon_m$, thus giving the polarizability of a two-layer coated particle in the host medium

$$\overline{\alpha}_i^{IEA} = R_2^3\frac{\overline{\varepsilon}_i-\varepsilon_m}{3\varepsilon_m+3L_{2i}(\overline{\varepsilon}_i-\varepsilon_m)}. \tag{15}$$

The second scheme looks more straightforward, but its accuracy is lower than the first. The physical explanation for this different accuracy is that IEA gives accurate cross sections for bare particles; therefore, we retain IEA accuracy in subsequent DEM applications by replacing polarizability through Eq. (6). If the coating thickness approaches zero, then both schemes give identical results.

Using Eq. (11), for the averaged over random particle orientations cross sections, we get[14]

$$\left\langle C_{ext}^{IEA}\right\rangle = \pi R_2^2\left\langle Q_{sca}^{IEA}\right\rangle = 4\pi\frac{1}{k'_m}\operatorname{Im}\left(k_m^2\frac{\overline{\alpha}_c^{IEA}+2\overline{\alpha}_a^{IEA}}{3}\right), \tag{16}$$

$$\left\langle C_{sca}^{IEA}\right\rangle = \pi R_2^2\left\langle Q_{sca}^{IEA}\right\rangle = \frac{8\pi}{3}\left|k_m\right|^4\frac{\left|\overline{\alpha}_c^{IEA}\right|^2+2\left|\overline{\alpha}_a^{IEA}\right|^2}{3}, \tag{17}$$

where $k'_m = \operatorname{Re}(k_m) = (2\pi/\lambda)\operatorname{Re}(n_m)$. Equations (16) and (17) are written for the general case of absorbing host. For bare particles without coating, Eqs. (16) and (17) reduce to Eqs. (55) and (56) of Ref.[14]. Note that in the EA approximation, Eq. (16) gives the absorption cross section rather than extinction; therefore, one has to calculate the extinction as a sum of (16) and (17).

However, such a procedure would be incorrect for IEA since IEA polarizability already considers the absorption contribution due to the imaginary term in the denominator of (4).

## 4. Modal expansion method for coated particles

### *4.1 Model*

The spheroidal shape is convenient for principal assessments of the particle shape effects. However, this form of particles practically never occurs.[26] In the gallery of typically synthesized gold nanoparticles (see Figure S10), elongated nanoparticles most often have the shape of rods with hemispherical or ellipsoidal ends. To illustrate the principles of the method, we will consider a gold rod with hemispherical ends of length $L_1$, diameter $d_1$, and aspect ratio $AR_1 \equiv h_1 = L_1 / d_1$. Here and below, indices 1 and 2 stand for the core and shell of a two-layer particle. For simplicity, we will consider only the excitation of the dominated plasmon resonance along the axis of symmetry of the rod; therefore, we omit the polarization index used in the previous section here. The shell's constant thickness and refractive index are equal to $s$ and $n_2 \equiv n_s = \sqrt{\varepsilon_2}$, respectively. The two-layered rod's total length, diameter, and aspect ratio are $L_2 = L_1 + 2s$, $d_2 = d_1 + 2s$, and $AR_2 \equiv h_2 = L_2 / d_2$, respectively. The core and total particle volumes are expressed as $V_i = L_i^3 \pi (3h_i - 1)/(12h_i^3)$ and are characterized by the radii of equivolume spheres $R_{iev} = \sqrt[3]{3V_i / 4\pi}$.

### *4.2 Polarizability of coated metal particles in MEM*

In MEM [10], the optical properties of nanoparticles are described by expansion of polarizability in terms of electrostatic eigenvalues[27]

$$\alpha(\lambda) = \frac{1}{4\pi} \sum_j V_j^m \left[ \frac{\varepsilon_m}{\varepsilon_1 - \varepsilon_m} - \frac{1}{\eta_j - 1} - A_j(x) \right]^{-1} . \tag{18}$$

where $x=\sqrt{\varepsilon_m} L_c / \lambda$, $L_c$ is a characteristic particle size (for example, the sphere diameter or rod length), $V_j^m$ is the modal volume that obeys the sum rule $\sum_j V_j^m = V$, $V$ is the total particle volume, $\eta_j$ is the eigenvalue of j-mode ( $j=1$ corresponds to the dipole eigenmode). In Eq. (18), the parameter $A_j(x)$ describes the retardation correction (also termed dynamic depolarization[28,29]) of polarizability due to finite particle size ($\sim x^2$), the radiative reaction correction ($\sim x^3$) (see Section 16.2 in Ref.[30]), and an additional term ($\sim x^4$)

$$A_j(x) = a_{j2} x^2 + i \frac{4\pi^2 V_j^m}{3L^3} x^3 + a_{j4} x^4 . \tag{19}$$

For a given particle model, parameters $\eta_j$, $V_j^m$, $a_{j2}$, $a_{j4}$ can be calculated by minimizing the deviations of the MEM solution from simulations using a suitable numerical method. Specifically, in Ref.[10] the particle shape was characterized by a generalized aspect ratio parameter (e.g., for rods, it equals the length/diameter ratio, while for disks, it equals the diameter/thickness ratio). Thus, after the minimization procedure, the MEM parameters $\eta_j$, $V_j^m$, $a_{j2}$, $a_{j4}$ are approximated only by simple analytical functions of the aspect ratio. Significantly, they do not depend on the particle size, composition, or environment. As a result, it is possible to find elementary still accurate analytical models to predict the electromagnetic response of variously shaped plasmonic nanoparticles. Owing to their simplicity, the MEM models can be easily applied to realistic experimental models that include polydisperse and polymorphic ensembles with any desk PC.

The generalization of MEM approaches to the case of particles with a dielectric or other coating is of undoubted theoretical and practical interest since, in practice, all particles are usually stabilized by ligands or enclosed in a composite shell. We plan to consider this problem in a separate publication; here, we will limit ourselves to a description of the general principle and its illustration using the practically important example of coated gold nanorods. Let us recall that all gold nanorods have a stabilizing CTAB layer on their surface with a thickness of about 3 nm.[22]

Fortunately, generalizing MEM to the case of multilayer particles does not require new calculations using numerical methods and new approximation of the MEM parameters $\eta_j$, $V_j^m$, $a_{j2}$, $a_{j4}$. It turns out that the problem can be easily solved using the previously developed dipole equivalence method (DEM), which has been applied to multilayer spheres[31] and spheroids.[13] Moreover, for simplicity, following Ref.[10] we will limit ourselves to only the contribution of the first mode since this approximation is entirely accurate for the case of longitudinal excitation of the rod.

To apply the DEM, we recast first Eq. (18) in the following equivalent form

$$\alpha_1 = R_{1ev}^3 \frac{\varepsilon_1 - \varepsilon_m}{3\varepsilon_m + 3L_{1eff}(h_1, \varepsilon_m)(\varepsilon_1 - \varepsilon_m)}, \tag{18a}$$

where we have introduced an "effective depolarization factor" which accounts for arbitrary particle shape together with the retardation and radiation-damping effects

$$L_{1eff}(h_1, \varepsilon_m) = \frac{1}{1 - \eta_1(h_1)} - A_1(h_1, \varepsilon_m), \tag{20}$$

$$A_1(h_1, \varepsilon_m) = a_{12}(h_1) x_{1m}^2 + i\frac{2}{9} X_{1m}^3 + a_{14}(h_1) x_{1m}^4, \tag{21}$$

$$X_{1m} = R_{1ev} k_m,\ x_{1m} = L_1 k_m / 2\pi,\ k_m = k\sqrt{\varepsilon_m} = (2\pi/\lambda)\sqrt{\varepsilon_m}, \tag{22}$$

where $R_{1ev} = \sqrt[3]{3V_1^m / 4\pi}$ is the equivolume sphere radius for the modal volume $V_1^m$, $L_1$ is the characteristic size of the core. In expression (21), we recognize the radiation damping term, which takes the well-known form $i\frac{2}{3} X_{1m}^3$ after multiplying by 3 in the denominator of Eq. (18a). Indeed, it reduces to $i\frac{2}{3}(k_m a)^3$ for spheres of a radius $a$. For spheroids with semiaxes $(a, a, c)$, we arrive at $i\frac{2}{3} h_1^{-2} \left(k_m c\right)^3$, which corresponds to the denominator in Eq. (4). Thus, Eq. (18a) exactly reproduces well-known results[8,29] for the radiation damping correction of electrostatic polarizability for spheres, spheroids, and other particles in terms of the equivolume radius parameter $X_{1m} = k_m R_{1ev}$. Similarly, the retardation correction term has a familiar quadratic behavior $\sim x_{1m}^2$, but it also involves the shape-dependent contribution $a_{12}(h_1)$, which differs from known size-dependent expansions of Mie spheres.[29] Finally, the last term in Eq. (21) accounts for $x_{1m}^4$ contribution with shape-dependent coefficient $a_{14}(h_1)$.

What was said above has been applied only to the metal core so far. Let us now turn to the case of coated particles. The form of expression (19) for polarizability has the same structure as the standard EA approximation (5), so we can apply the DEM scheme (Figure 1) with the only difference that in the first step, we use the bulk metal dielectric function $\varepsilon_1$ or its counterpart obtained taking into account the limitation of the electron free path due to surface scattering.[32] In the second step, we place the rod in a medium with a dielectric constant $\varepsilon_2$; from the equality of the dipole moments of particles in panels step 2 and step 3 (Figure 1), we obtain the following equation for the average dielectric constant $\overline{\varepsilon}_2$ of a particle in step 3:

$$R_{1ev}^3 \frac{\varepsilon_1 - \varepsilon_2}{3\varepsilon_2 + 3L_{1ff}(h_1, \varepsilon_2)(\varepsilon_1 - \varepsilon_2)} = R_{2ev}^3 \frac{\overline{\varepsilon}_2 - \varepsilon_2}{3\varepsilon_2 + 3L_{2ff}(h_2, \varepsilon_2)(\overline{\varepsilon}_2 - \varepsilon_2)}, \tag{23}$$

where

$$L_{ieff}(h_i, \varepsilon_2) = \frac{1}{1 - \eta_1(h_i)} - A_1(h_i, \varepsilon_2), \quad i = 1,2, \tag{24}$$

$$A_1(h_i, \varepsilon_2) = a_{12}(h_i) x_{i2}^2 + i\frac{2}{9} X_{i2}^3 + a_{14}(h_i) x_{i2}^4, \quad i = 1,2, \tag{25}$$

$$x_{i2} = L_i k_2 / 2\pi, \quad X_{i2} = R_{iev} k_2, \quad i = 1,2, \tag{26}$$

where indices $i = 1,2$ stand for the core and shell, respectively and $R_{2ev} = \sqrt[3]{3V_2^m / 4\pi}$ is the equivolume sphere radius for the modal volume $V_2^m$. From Eq. (23) we find

$$\overline{\varepsilon}_2 = \varepsilon_2 \left( 1 + \frac{D_1}{1 - L_{2eff}(\varepsilon_2, h_2) D_1} \right), \tag{27}$$

with

$$D_1 = f_{12} \frac{\varepsilon_1 - \varepsilon_2}{\varepsilon_2 + L_{1eff}(h_1, \varepsilon_2)(\varepsilon_1 - \varepsilon_2)}, \tag{28}$$

where $f_{12} = (R_{1ev} / R_{2ev})^3$ is the core volume fraction. Finally, for the resulted MEM polarizability of a coated plasmonic particle, we get

$$\bar{\alpha}_2^{MEM} = R_{2ev}^3 \frac{\bar{\varepsilon}_2 - \varepsilon_m}{3\varepsilon_m + 3L_{2eff}(h_2, \varepsilon_m)(\bar{\varepsilon}_2 - \varepsilon_m)}, \quad (29)$$

$$L_{2eff}(h_2, \varepsilon_m) = \frac{1}{1 - \eta_1(h_2)} - A_1(h_2, \varepsilon_m),, \quad (30)$$

$$A_1(h_2, \varepsilon_m) = a_{12}(h_2) x_{2m}^2 + i\frac{2}{9} X_{2m}^3 + a_{14}(h_2) x_{2m}^4, \quad (31)$$

$$x_{2m} = L_2 k_m / 2\pi, \quad X_{2m} = R_{2ev} k_m. \quad (32)$$

where $L_2$ is the characteristic size of the shell. The above consideration can be easily applied to other polarizations of the incident light, provided that the modal parameters $\eta_j, a_{2j}, a_{4j}$ have been calculated for desired modes and polarizations. Note that the modal volumes $V_i^m$ for the bare ($i = 1$) and coated ($i = 2$) rods can be approximated as $V_i^m = 0.896 V_i$.[10]

*4.3 MEM polarizability of metal particles with a multilayered coating*

The MEM+DEM scheme can be easily generalized to multilayered particles. To illustrate the main idea, we consider an arbitrary metal particle with two dielectric layers. In line with[10] and consideration in Sections 2.1-2.2, the metal particle is characterized by a length $L_1$ aspect ratio $h_1$, and dielectric function $\varepsilon_1$. The particle is coated by two dielectric layers with thicknesses $s_2$, $s_3$ and dielectric functions $\varepsilon_2$, $\varepsilon_3$. The characteristic lengths and aspect ratios of coated particles are $L_i$ and $h_i$, where indices $i = 2,3$ stand for the first and second dielectric layers, respectively.

After applying MEM+DEM scheme to the particle with the first dielectric layer, we arrive at the MEM polarizability (29) with the averaged dielectric function $\bar{\varepsilon}_2$. To calculate the MEM polarizability of a three-layered particle, we place the two-layered particle in a medium with a dielectric constant $\varepsilon_3$ and introduce an imaginary surface coinciding with the third dielectric layer. In other words, we consider steps 2 and 3 in Figure 1 where dielectric functions $\tilde{\varepsilon}_{1i}$, $\bar{\varepsilon}_i$, and $\varepsilon_2$ are replaced with $\overline{\varepsilon}_2$, $\overline{\varepsilon}_3$, and $\varepsilon_3$, respectively. Then, in the spirit of DEM, from the equality of the dipole moments of particles in steps 2 and 3, we obtain the following equation for the average dielectric constant $\overline{\varepsilon}_3$ of the three-layered particle in step 3

$$\overline{\varepsilon}_3 = \varepsilon_3 \left( 1 + \frac{D_3}{1 - L_{3eff}(h_3, \varepsilon_3) D_3} \right), \tag{33}$$

with

$$D_3 = f_{23} \frac{\overline{\varepsilon}_2 - \varepsilon_3}{\varepsilon_3 + L_{2eff}(h_2, \varepsilon_3)(\overline{\varepsilon}_2 - \varepsilon_3)}, \tag{34}$$

where $f_{23} = (R_{2ev} / R_{3ev})^3$ is the volume fraction of the two-layered particle, $R_{3ev} = \sqrt[3]{3V_3^m / 4\pi}$ is the equivolume sphere radius for the modal volume $V_3^m$. Finally, for the resulting MEM polarizability of the plasmonic particle with two dielectric layers, we get

$$\overline{\alpha}_3^{MEM} = R_{3ev}^3 \frac{\overline{\varepsilon}_3 - \varepsilon_m}{3\varepsilon_m + 3L_{3eff}(h_3, \varepsilon_m)(\overline{\varepsilon}_3 - \varepsilon_m)}, \tag{35}$$

$$L_{3eff}(h_3, \varepsilon_m) = \frac{1}{1 - \eta_1(h_3)} - A_1(h_3, \varepsilon_m),, \tag{36}$$

$$A_1(h_3, \varepsilon_m) = a_{12}(h_3) x_{3m}^2 + i \frac{2}{9} X_{3m}^3 + a_{14}(h_3) x_{3m}^4, \tag{37}$$

$$x_{3m} = L_3 k_m / 2\pi, \quad X_{3m} = R_{3ev} k_m. \tag{38}$$

where $L_3$ is the characteristic size of the third layer. In Eqs. (33) and (34), the effective depolarization factors $L_{3eff}(h_3, \varepsilon_3)$ and $L_{2eff}(h_2, \varepsilon_3)$ are calculated by Eqs. (35)-(38) where $\varepsilon_m$ should be replaced with $\varepsilon_3$ and $h_3$ with $h_2$ (for $L_{2eff}(h_2, \varepsilon_3)$). Numerical illustration of the above scheme will be considered elsewhere.

## 5. Results and discussion.

### 5.1. IEA for coated particles

#### *5.1.1. Prolate gold spheroids*

Figure 1 shows the extinction spectra of randomly oriented elongated gold spheroids with a diameter of 20 nm and a length from 40 to 200 nm. Here and below, we focus on the dominant peak caused by the plasmon resonance along the particle's major axis.

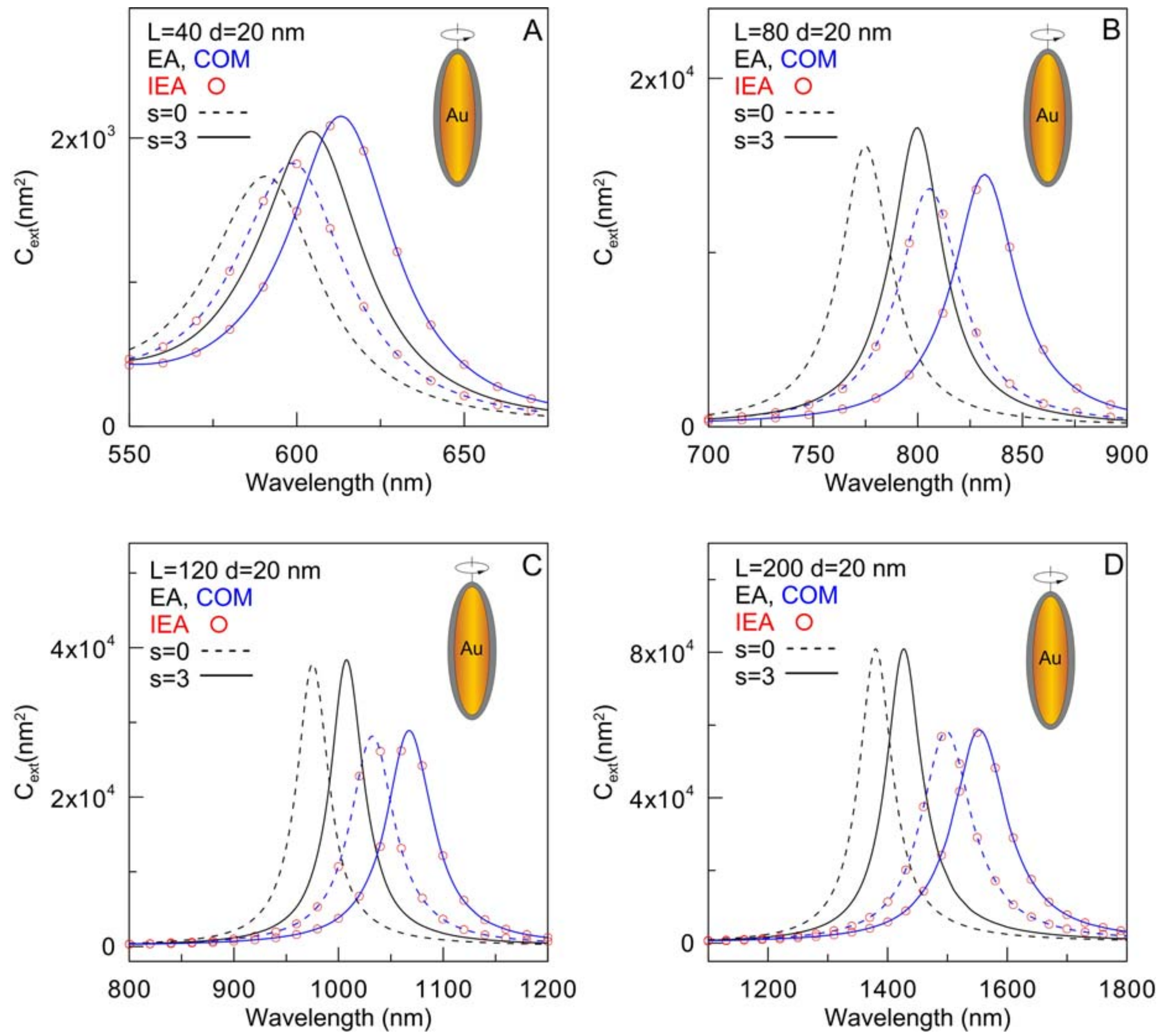

**Figure 2**. Extinction spectra of randomly oriented Au bare and coated randomly oriented spheroids were calculated using EA (black), IEA (red circles), and COM (blue). The particle diameter $d = 2a_1$ is 20 nm, and their length $L = 2c_1$ is 40 (A), 80 (B), 120 (C), and 200 nm. The coating thickness is 0 (dashed lines) and 3 nm (solid line); the coating refractive index is 1.5.

First, we note that the usual EA approximation gives inaccurate spectra, starting from a particle length of 40 nm. On the contrary, the IEA approximation is in excellent agreement with calculations of extinction spectra using COMSOL. Coating nanoparticles with a dielectric shell leads to the well-known red shift of the main plasmon peak.[33,34] For the smallest particles with a length of 40 nm, the dielectric coating leads to a slight increase in the resonance amplitude, and for particles with a length of 200 nm, the resonance amplitude of the 3 nm coating does not change. For scattering cross sections, the spectra are largely similar to those in Figure 2 (see Figure S1, SI). To verify the accuracy of simulated spectra, we performed benchmark calculations with the T-matrix method (TM), SVM, and COMSOL (Figure S2). It follows from Figure S2 that all three methods are in excellent agreement even for long spheroids with the length $L = 160$ nm and aspect ratio 8. For large oblate spheroids (Figures S2 C and D), TM and COM spectra also coincide

When the coating thickness is increased by an order of magnitude (Figure 3), then IEA still gives perfect accuracy in the position of the longitudinal resonance of extinction and scattering but slightly underestimates its amplitude. More noticeable differences are observed in the short-wavelength region of transverse resonance (see insets in Figure 3). The physical reason for this inaccuracy of IEA may be an inadequate representation of the polarizability for a metal particle with a highly non-uniform confocal dielectric coating (equation (2)) by an average polarizability of a homogeneous particle.

In most biomedical applications, the diameter and length of gold nanorods do not exceed 20 and 120 nm, respectively. Thus, we conclude that a simple IEA approximation can be used to model the extinction and scattering spectra of dielectric-coated elongated gold particles.

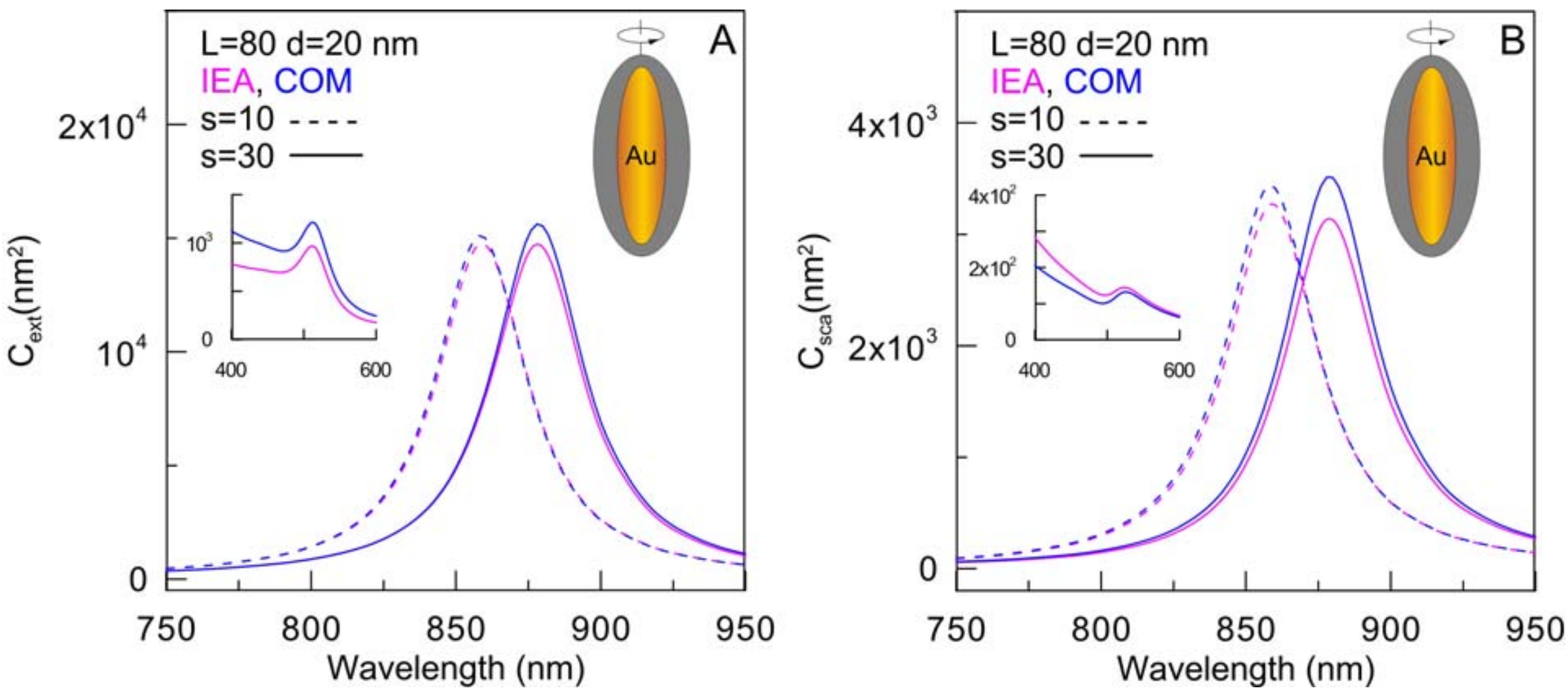

**Figure 3**. Extinction (A) and scattering (B) spectra for a gold core $L = 2c_1 = 80$ nm, $d = 2a_1 = 20$ nm and the shell thickness 10 (dashed lines) and 30 nm (solid lines).

*5.1.2. Oblate gold spheroids*

For calculations, the length of the symmetry axis of the spheroids (in other words, the particle thickness) was fixed and equal to 20 nm, and the diameter varied from 40 to 120 nm. As for elongated particles, the EA approximation is no longer acceptable for particles with diameters of 40 and 80 nm. At the same time, the cross sections calculated from IEA and COMSOL are in good agreement with each other (Figures 4 A, B). However, for oblate particles, the volume depends quadratically on the diameter, so for $L = 2c_1 = 20$ nm, $d = 2a_1 = 120$ nm particles, the equivolume sphere radius is $R_{1ev} = 33$ nm, while for the largest elongated particles in Figure 2 $R_{1ev} = 21.5$ nm. Consequently, restrictions on the range of IEA applicability for the diameter of oblate particles occur earlier than for the length of prolate spheroids. Differences between the IEA and COMSOL spectra appear for diameters 100 and 120 nm (panels C and D). The spectra of scattering cross sections (Figure S3, SI) are largely similar to those in Figure 4.

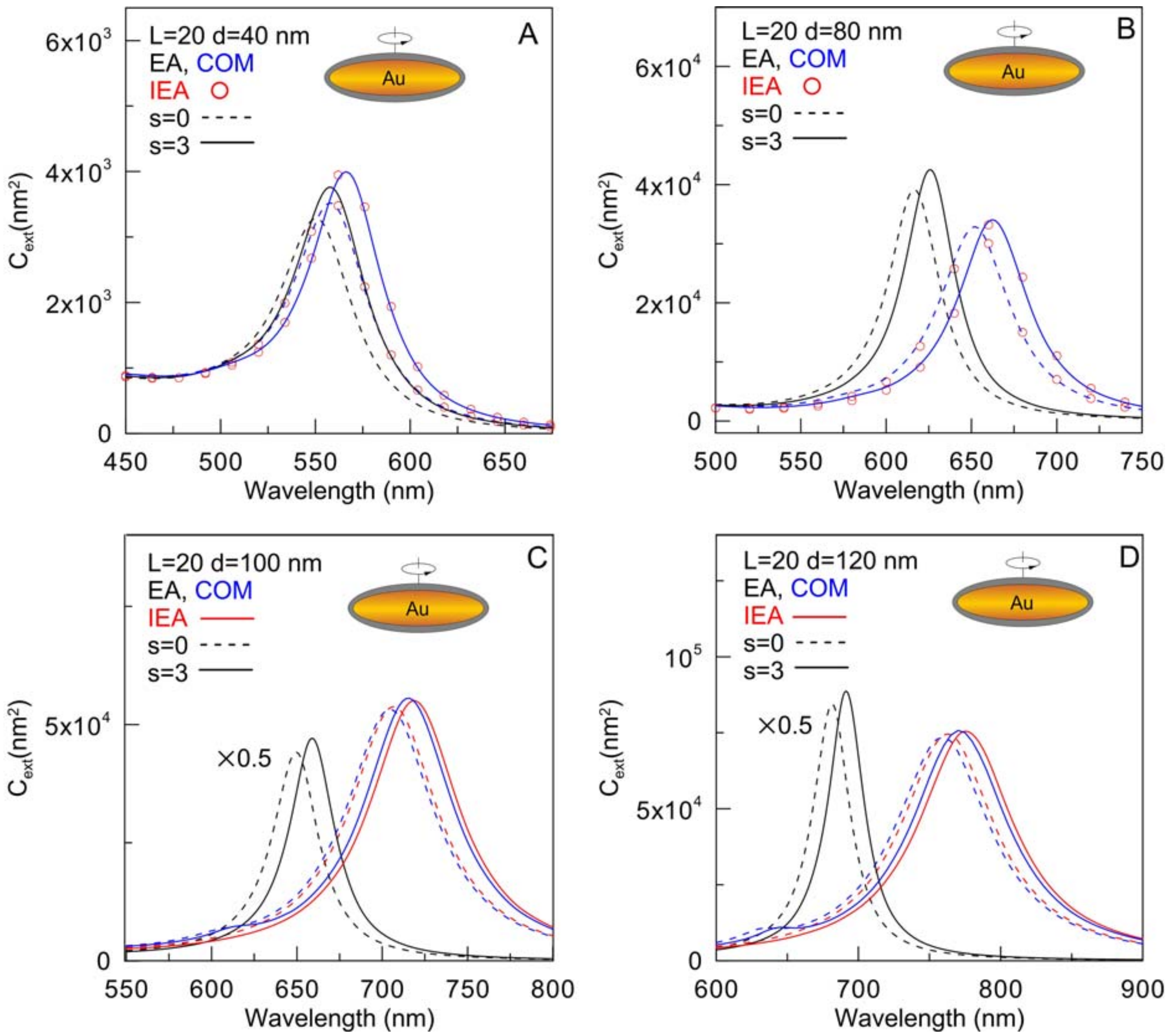

**Figure 4**. Extinction spectra of randomly oriented Au bare and coated randomly oriented oblate spheroids were calculated using EA (black), IEA (red circles), SVM (A, B), and COM (C, D) (blue). The particle thickness is $L = 2c_1 = 20$ nm, and their diameter is 40 (A, 80 (B), 100 (C), and 120 nm. The coating thickness is 0 (dashed lines) and 3 nm (solid line); the coating refractive index is 1.5. Note that EA spectra in panels (C, D) were multiplied by 0.5 for convenience.

Starting from a diameter of 100 nm, a quadrupole shoulder appears in the extinction spectra (about 600 nm in panel C and about 630 nm in panel D), which the IEA approximation cannot reproduce. For example, for oblate bilayer particles with sizes 20-120-3 nm (*L*-*d*-*s*; Figure S4 A, SI), the short-wavelength part of the extinction spectrum according to IEA is clearly underestimated compared to COMSOL data. However, the IEA approximation correctly describes the main plasmon peak's position and amplitude in agreement with COMSOL calculations (Figure S4-B, SI). Note that the calculations using our SVM program for uncoated gold particles with a diameter of 100 nm or more differed from those using the TM and COMSOL. Comparison with TM and COMSOL calculations showed (see Figures S2 C, D) that the differences obtained are apparently due to inaccurate data obtained using available SVM.

The effect of a thick shell (10 and 30 nm) was assessed for a core with $L = 2c_1 = 20$ nm, $d = 2a_1 = 80$ nm (Figure 5). In contrast to the case of a thin shell (Figure 4B), for thick shells, the IEA approximation slightly underestimates the amplitude of the main plasmon peak, but its position is predicted quite accurately.

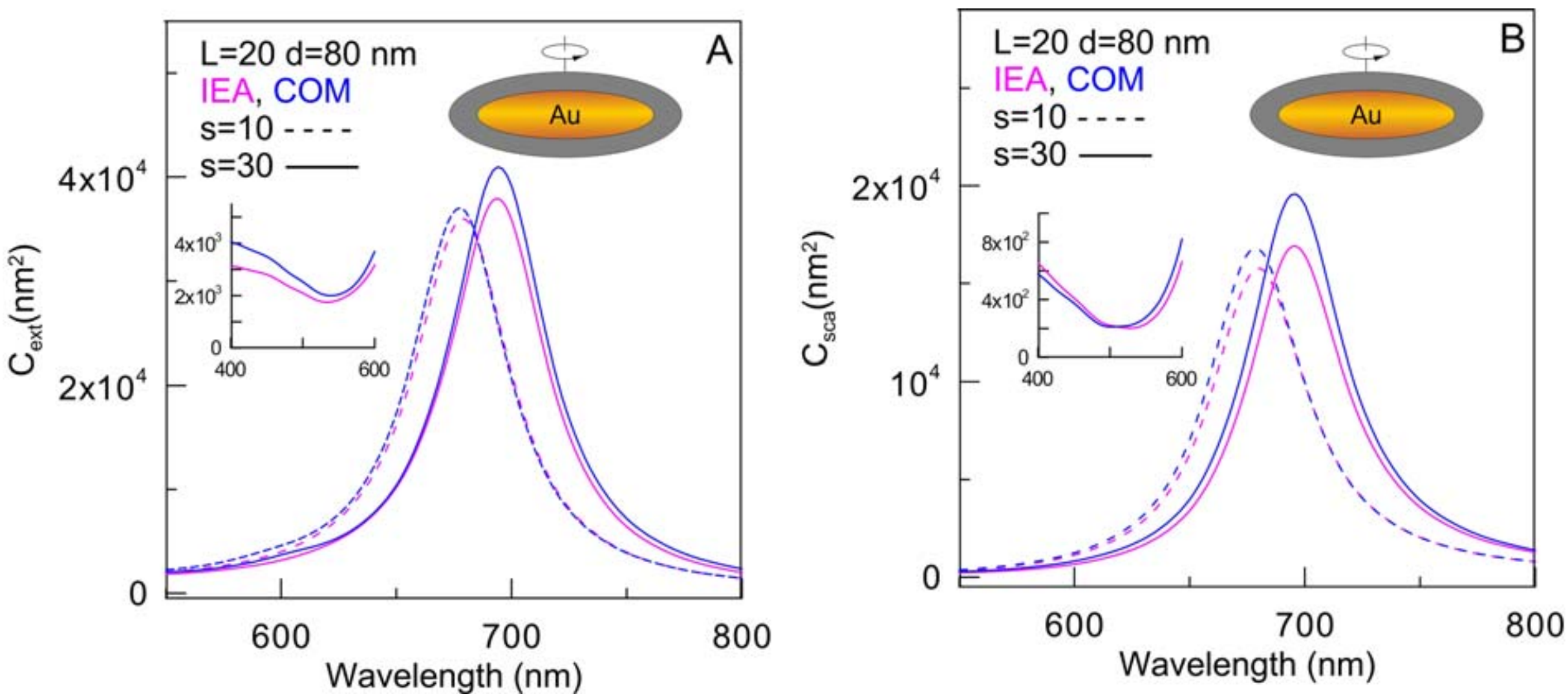

**Figure 5**. Extinction (A) and scattering (B) spectra for an oblate gold core $L = 20$ nm, $d = 80$ nm and the shell thickness 10 and 30 nm.

Thus, for the case of oblate gold particles, the IEA approximation can be used to simulate the extinction and scattering spectra of elongated gold particles with a dielectric coating if their diameter does not exceed 120 nm with a thickness of about 20 nm.

*5.1.3. Prolate silver spheroids*

Figure 6 shows extinction spectra calculated for the same parameters as in Figure 2 but for silver particles. As in the case of gold particles, the EA approximation cannot be used for either metal particles or coated particles, while the IEA approximation in the region of the main plasmonic dipole peak agrees well with the COMSOL and SVM calculations. However, in the short-wavelength region, IEA works only for small particles (Figure S5 A, SI). For particles with dimensions of 100x20 nm, there is a slight underestimation of the amplitude of the minor peak (Figure S6 A, SI).

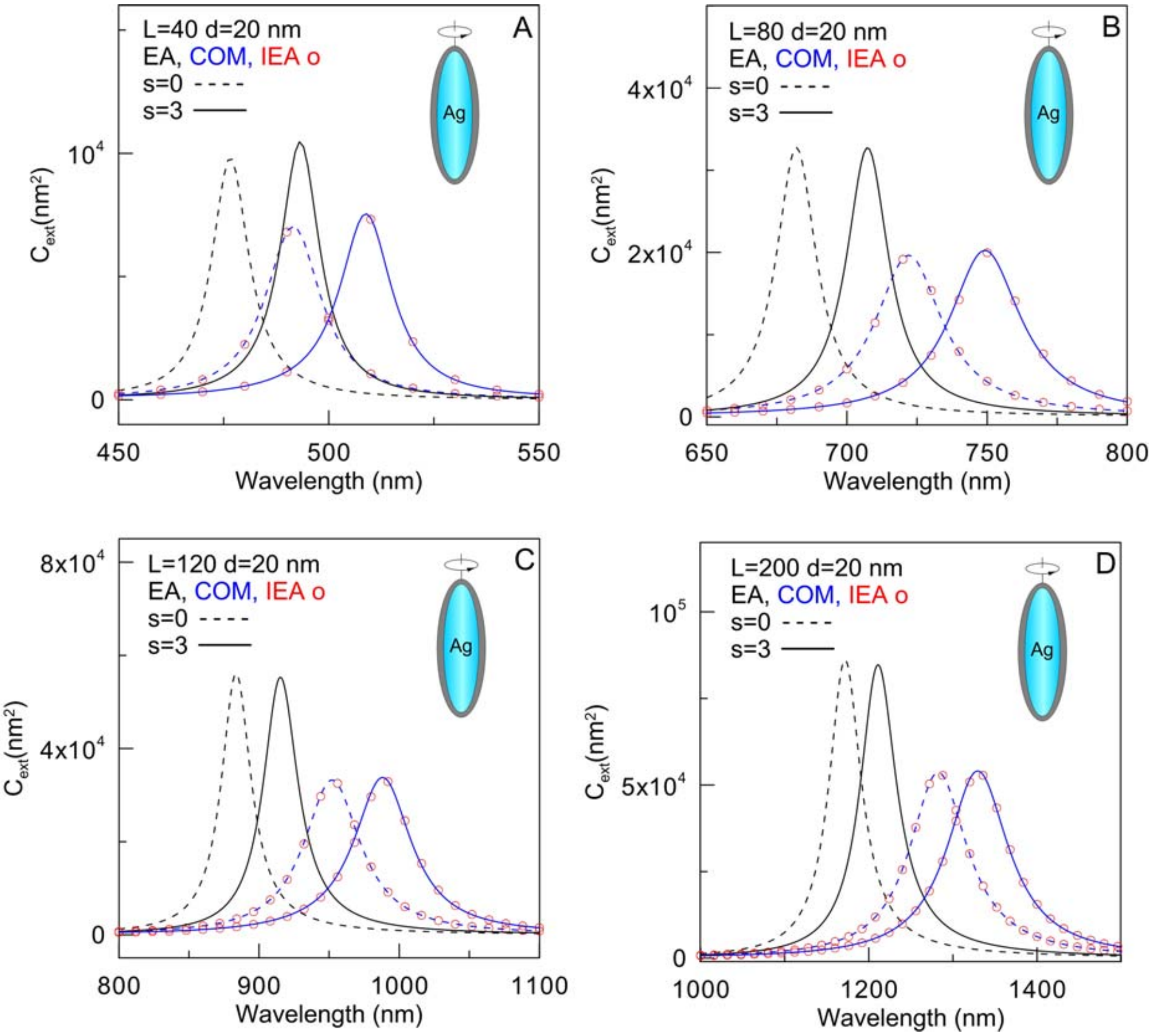

**Figure 6**. Extinction spectra of randomly oriented Ag bare and coated randomly oriented spheroids were calculated using EA (black), IEA (red circles), and COM (blue). The particle diameter is 20 nm, and their length is 40 (A), 80 (B), 120 (C), and 200 nm. The coating thickness is 0 (dashed lines) and 3 nm (solid line); the coating refractive index is 1.5.

With increasing coating thickness to 10 and 30 nm, there is a slight underestimation of the amplitude of the plasmon peak for the extinction and scattering cross sections (Figure 7), but its position is well predicted. Thus, the simple IEA approximation can be used to model the extinction and scattering spectra of dielectric-coated silver elongated particles since particles with a diameter greater than 20 nm and a length greater than 200 nm are not used in biomedical applications.

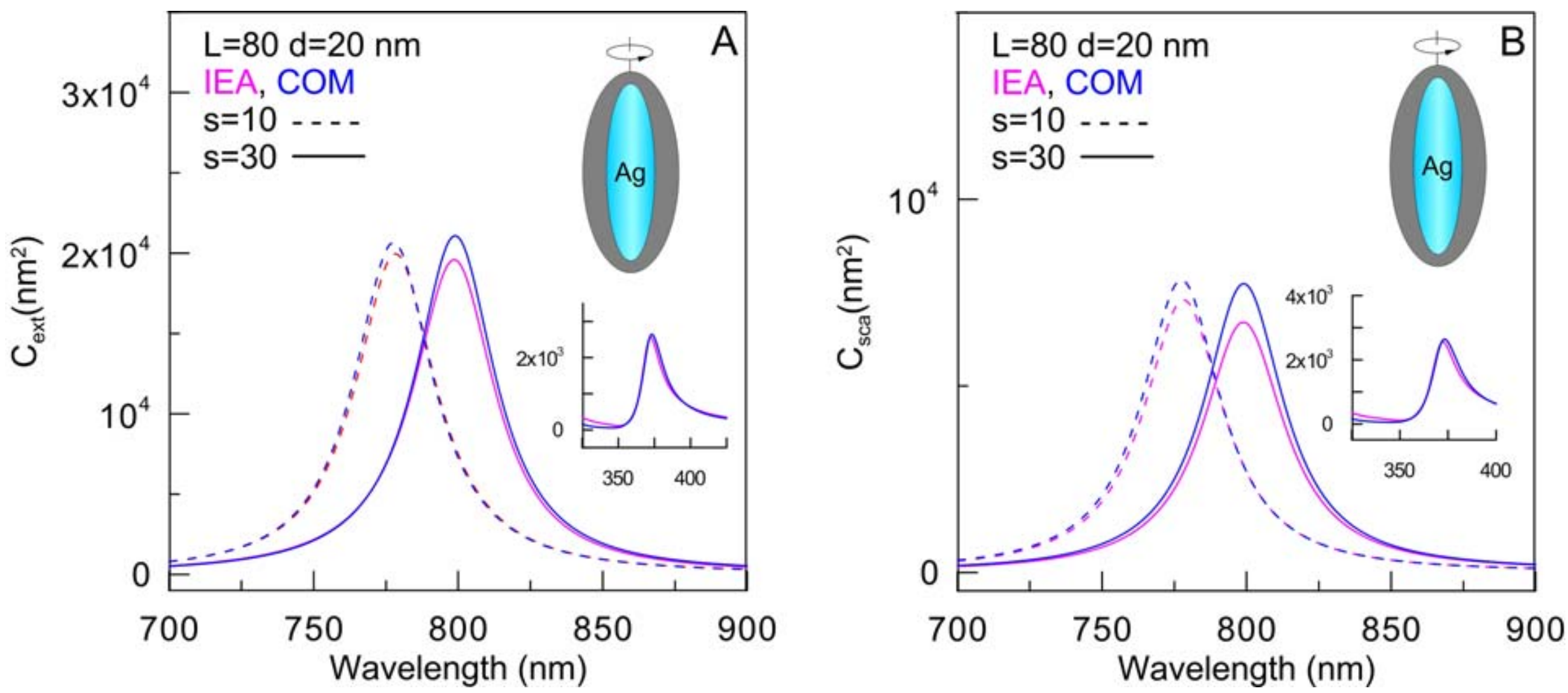

**Figure 7**. Extinction (A) and scattering (B) spectra for a silver core $L = 2c_1 = 80$ nm, $d = 2a_1 = 20$ nm and the shell thickness 10 (dashed lines) and 30 nm (solid lines).

### *5.1.4. Oblate silver spheroids*

In the case of large-volume oblate silver particles, the SVM program produced inaccurate data, so we used COMSOL to calculate the extinction and scattering spectra of the metallic and coated particles for comparison with the IEA spectra. The accuracy of COMSOL calculations for metal particles was verified using the T-matrix method (Figure S7).

Figure 8 shows the extinction spectra of silver nanoparticles in water without coating and with a 3-nm coating in the region of the main dipole peak. Only for the smallest particles $10 \times 40$ nm do COMSOL spectra coincide with IEA calculations. For particles with a thickness of 20 nm, the IEA spectra are red-shifted compared to the COMSOL ones. In addition, minor multipole peaks appear in the spectra for particles with a diameter greater than 100 nm, which the IEA approximation cannot reproduce (see Figure S8, SI). As the shell thickness increases to 30 nm, the red shift of the spectra is about 60 nm (Figure 9). Note that for such thick shells, the IEA approximation describes the spectral shift well but slightly underestimates the peak amplitude.

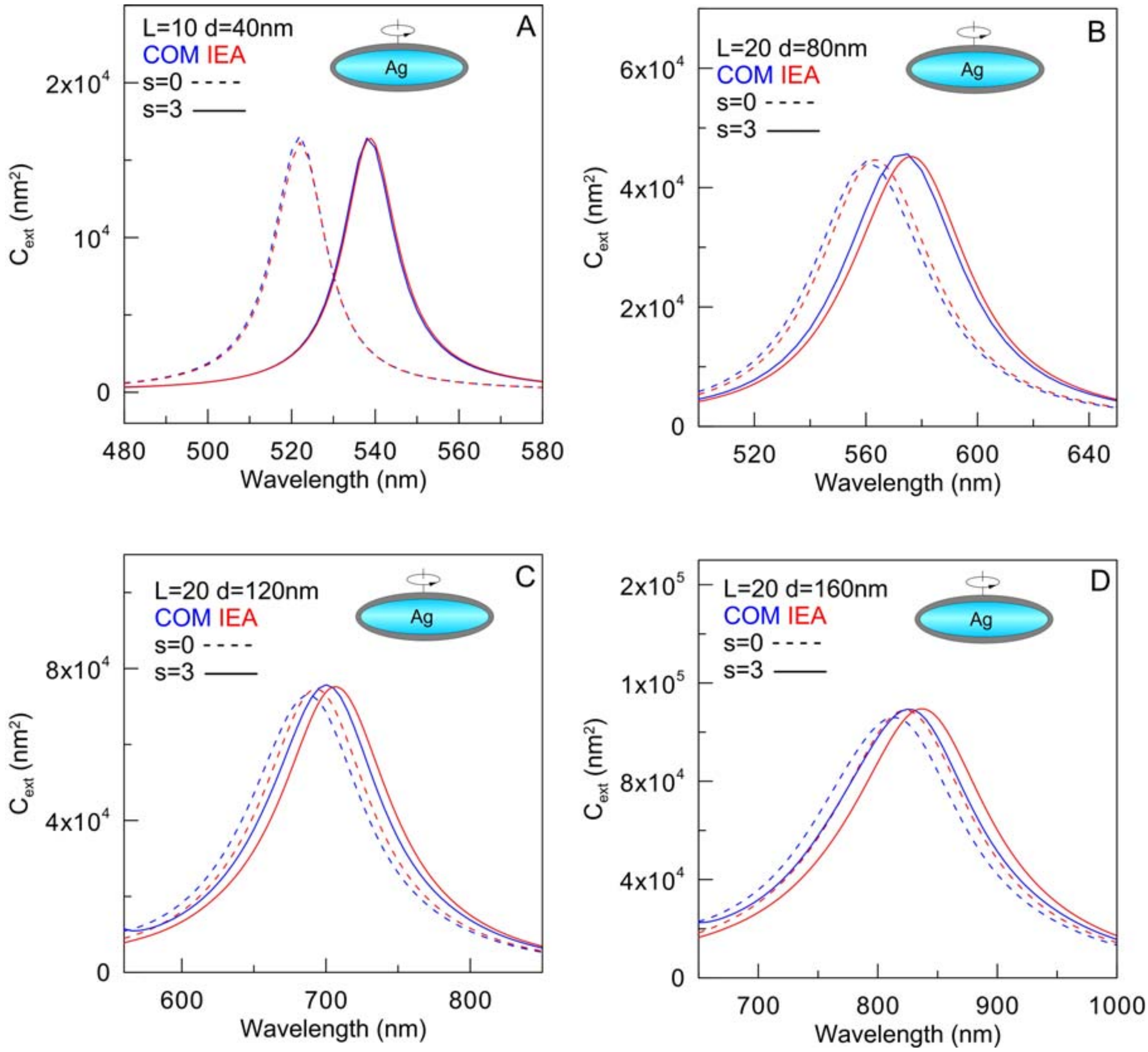

**Figure 8**. Extinction spectra of randomly oriented Ag bare and coated randomly oriented spheroids were calculated using COMSOL (blue) and IEA (red). The particle diameter is 20 nm, and their length is 40 (A), 80 (B), 120 (C), and 200 nm. The coating thickness is 0 (dashed lines) and 3 nm (solid line); the coating refractive index is 1.5.

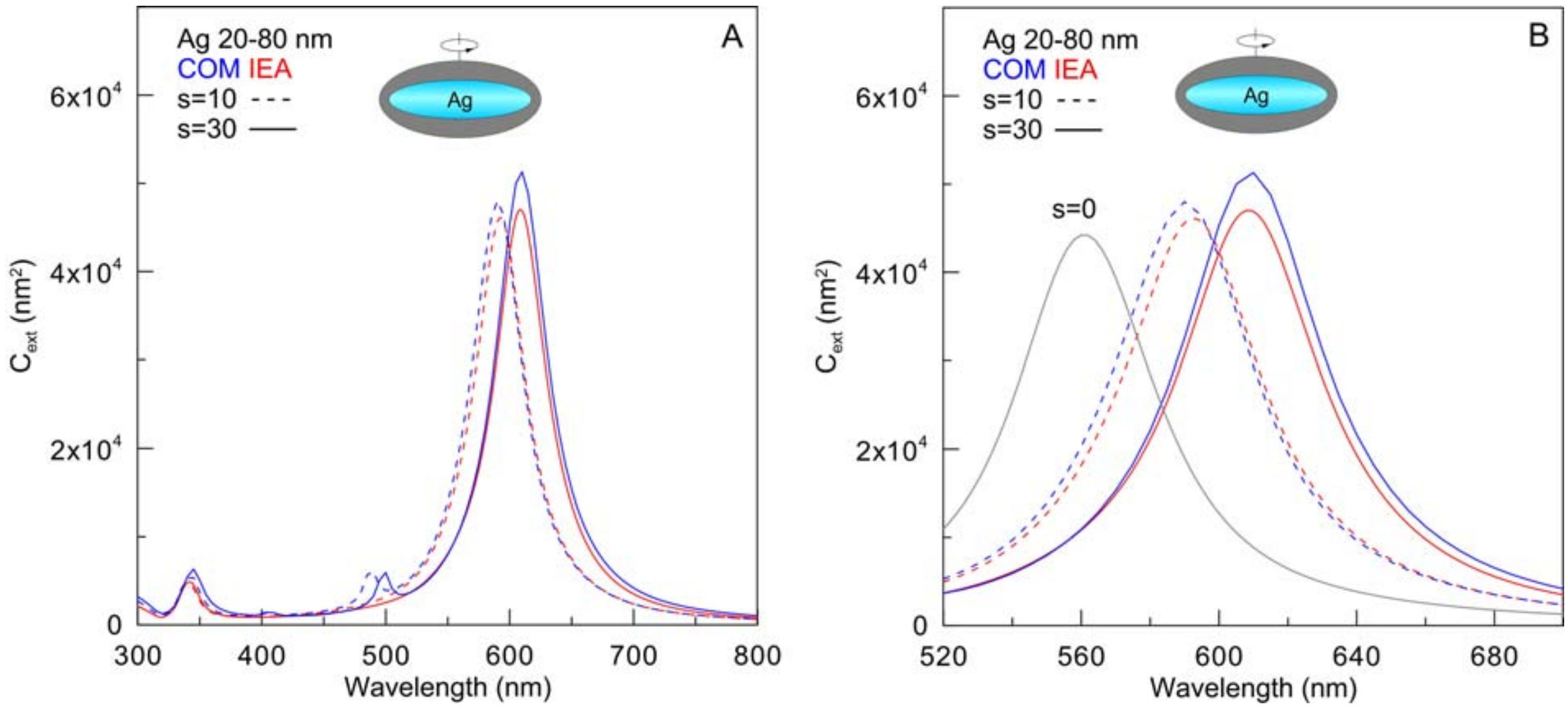

**Figure** 9. Extinction spectra of randomly oriented oblate Ag bare and coated spheroids were calculated using COMSOL (blue) and IEA (red). Panel B shows an enlarged portion of spectra near the main dipole peak. The particle diameter is 80 nm, and their thickness is 20 nm. The coating thickness is 10 (dashed lines) and 30 nm (solid line); the refractive index of the coating is 1.5. The black line in panel B shows the extinction spectrum for bare particles.

In practical studies, particles are often placed in a porous silicate shell with a refractive index of about 1.5.[35] This procedure ensures the stability of plasmonic properties in the bloodstream and during aggregation on a biotarget. In addition, the silicate shell provides a convenient platform for functionalization. The data in Figure 9 shows that the simple IEA approximation can be a convenient alternative to numerical methods when modeling the spectra of nanoparticles with a thick shell.

### 5.2. MEM for coated gold nanorods

For simulations, we chose the nanorod diameter of 15 nm, which is typical for experimental conditions.[26,36] Figures 10A and 10B compare extinction and scattering spectra computed for AuNRs with an aspect ratio from 2 to 6 by COMSOL and MEM (note that COMSOL and T-matrix data coincide here). The MEM parameters as a function of the aspect ratio were taken from.[10] In agreement with,[10] we obtained perfect MEM accuracy for extinction spectra within main plasmon resonances. However, as expected, MEM does not reproduce multipole plasmon modes within the 500-700 nm spectral band, which are seen in plots for aspect ratios above 4 (Figure 10A). The same conclusion is also proper for scattering spectra (Figure 10B). For our goals, the most important is excellent accuracy of MEM for AuNRs with thin coating thickness ranging from 3 to 30 nm and with a typical dielectric refractive index $n_s = 1.5$ like that for CTAB (Figures 10 C and D). Of course, such accuracy is observed near the main plasmonic resonance band, as MEM cannot reproduce the multipole minor peaks.

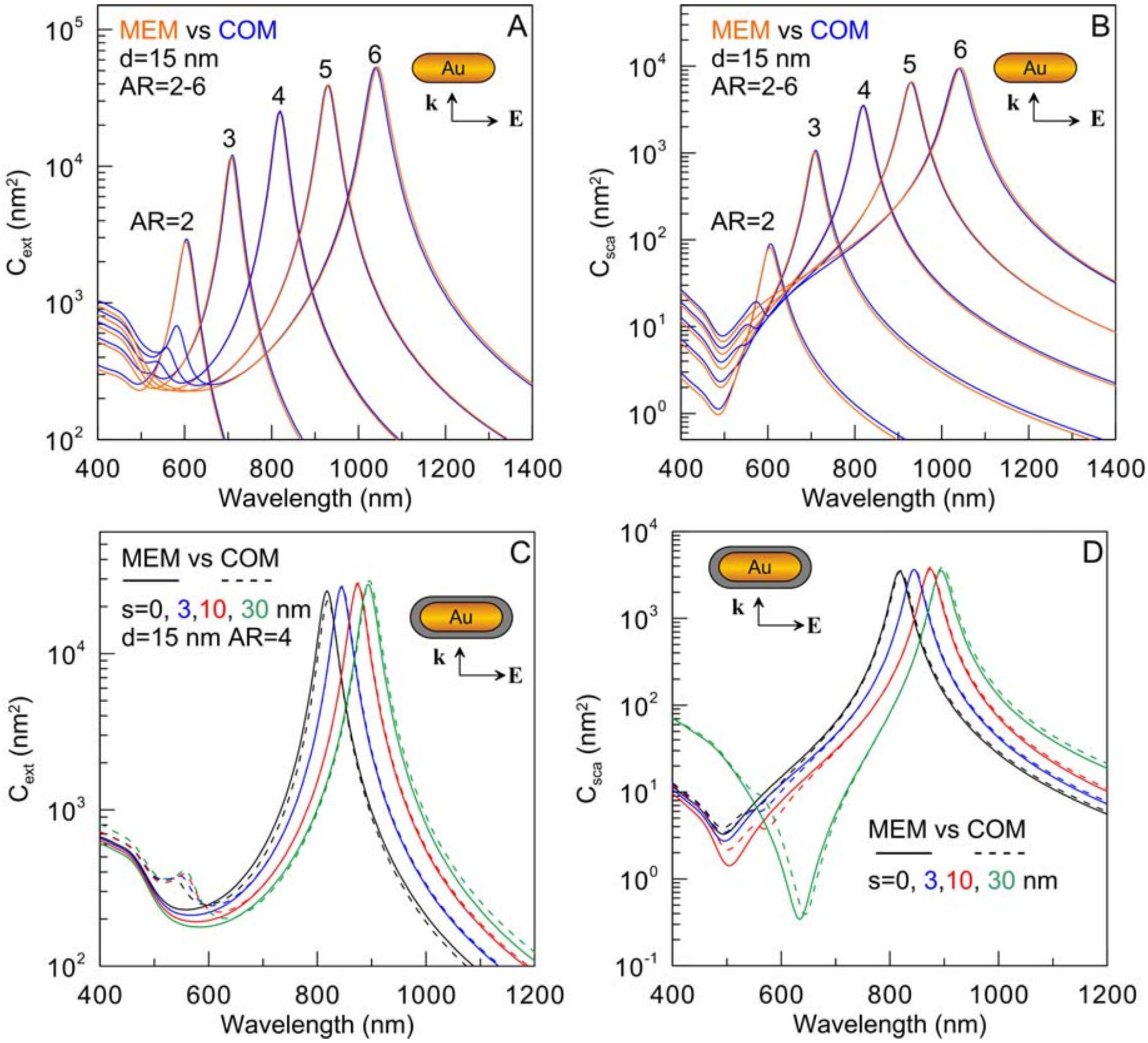

**Figure 10**. Extinction (A, C) and scattering (B, D) spectra of bare (A, B) and coated (C, D) gold nanorods with a diameter of 15 nm, the aspect ratio ranging from 2 to 6, and a coating thickness from 3 to 30 nm. Comparison of MEM *vs* COMSOL calculations (here, COM and TM produce

identical spectra). The nanorods are excited along the symmetry axis. Note that MEM spectra do not reproduce the quadrupole peaks near 550 nm. For a maximal 30-nm coating, green spectra in panel D demonstrate strong short-wavelength scattering, and a red shift of scattering minima near 630 nm is caused by the dielectric shell.

To summarize, we expect the great potential of the generalized MEM as applied to coated plasmonic particles. More detailed simulations for coated prolate and oblate Au and Ag particles of various shapes will be presented elsewhere.

## Conclusions

We have expanded IEA and MEM approximations previously developed for bare plasmonic nanoparticles to the coated ones. Remarkably, our extension of MEM does not require any additional numerical simulations to obtain the MEM mode expansion functions. In fact, we used previously reported data for bare nanoparticles.[10] This significant simplification has been achieved by applying the dipole equivalence method[25] to the metal-particle versions of IEM and MEM.

Comparing analytical model data with electromagnetic numerical simulations by COMSOL, we have found excellent accuracy of the developed extended IEA and MEM version in the prediction of dominant plasmonic extinction and scattering peaks. Thus, the developed analytical models can find valuable applications in modeling various plasmonic properties of coated particles, including absorption and scattering spectra, plasmon quantum yield, SERS enhancement, and metal-enhanced fluorescence.

## Acknowledgments

This research was supported by the Russian Science Foundation (project no. 24-22-00017).

**Supporting Information**. The Supporting Information is available free of charge on the ACS Publication website at DOI: XXX. Section S1: Explicit formulas for IEA functions. Section S2:

Figure S1. Scattering spectra for bare and coated prolate randomly oriented Au spheroids calculated by EA, IEA, and COM; Figure S2. Extinction spectra of randomly oriented prolate and oblate Au spheroids calculated by TM, SVM, and COMSOL; Figure S3. Scattering spectra for bare and coated oblate randomly oriented Au spheroids calculated by EA, SVM, TM, and IEA; Figure S4. Extinction and scattering spectra for bare and coated oblate randomly oriented Au spheroids calculated using COMSOL and IEA; Figure S5-S7. Extinction spectra for bare prolate and oblate randomly oriented Ag spheroids calculated by TM, SVM, and IEA; Figure S8. Extinction spectra of randomly oriented Ag bare and coated randomly oriented spheroids calculated using COMSOL and IEA; Section S3. Optical constants of bulk silver and water. Section S4. Gallery of plasmonic nanoparticles fabricated at IBPPM RAS**.**

**References and notes**

(1) Thanh, N. T. K.; Green L. A. W. Functionalisation of Nanoparticles for Biomedical Applications. *Nano Today* **2010**, *5*, 213–230.

(2) Sapsford, K. E.; Algar, W. R.; Berti, L.; Gemmill, K. B.; Casey, B. J.; Oh, E.; Stewart, M. H.; Medintz, I. L. Functionalizing Nanoparticles with Biological Molecules: Developing Chemistries that Facilitate Nanotechnology. *Chem. Rev.* **2013**, *113*, 1904–2074.

(3) Dykman, L. A.; Khlebtsov, N. G. Immunological Properties of Gold Nanoparticles. *Chem. Sci*. **2017**, *8*, 1719–1735.

(4) Bansal, S. A.; Kumar, V.; Karimi, J.; Singh, A.P.; Kumar, S. Role of Gold Nanoparticles in Advanced Biomedical Applications. *Nanoscale Adv*. **2020**, *2*, 3764–3787.

(5) Zheng, J.; Cheng, X.; Zhang, H.; Bai, X.; Ai, R.; Shao, L.; Wang, J. Gold Nanorods: The Most Versatile Plasmonic Nanoparticles. *Chem. Rev*. **2021**, *121*, 13342–13453.

(6) Myroshnychenko, V.; Rodríguez-Fernández, J.; Pastoriza-Santos, I.; Funston, A. M.; Novo, C.; Mulvaney, P.; Liz-Marzán, L. M.; García de Abajo, F. J. Modelling the Optical Response of Gold Nanoparticles. *Chem. Soc. Rev*. **2008**, *37*, 1792–1805.

(7) Bohren, C. F.; Huffman, D. R. *Absorption and Scattering of Light by Small Particles*, Wiley: New York, 1983.

(8) Majic, M.; Pratley, L.; Schebarchov, D.; Somerville, W. R. C.; Auguie, B.; Le Ru, E. C. Approximate T matrix and Optical Properties of Spheroidal Particles to Third Order with Respect to Size Parameter. *Phys. Rev. A* **2019**, *99*, 013853.

(9) Khlebtsov, N. G.; Le Ru, E. C. Analytical Solutions for the Surface and Orientation Averaged SERS Enhancement Factor of Small Plasmonic Particles. *J. Raman Spectrosc*. **2021**, *52*, 285–295.

(10) Yu, R.; Liz-Marzán, L. M.; García de Abajo, F. J. Universal Analytical Modeling of Plasmonic Nanoparticles. *Chem. Soc. Rev*. **2017**, *46*, 6710–6724.

(11) Farafonov, V. G.; Voshchinnikov, N. V.; Somsikov, V. V. Light Scattering by a Core-Mantle Spheroidal Particle. *Appl. Opt*. **1996**, *35*, 5412–5426.

(12) Ciracì, C.; Urzhumov, Y.; Smith, D. R. Far-Field Analysis of Axially Symmetric Three-Dimensional Directional Cloaks. *Opt. Express* **2013**, *21*, 9397–9406.

(13) Gladyshev, S.; Pashina, O.; Proskurin, A.; Nikolaeva, A.; Sadrieva, Z. Bogdanov, A.; Petrov, M.; Frizyuk, K. Fast Simulation of Light Scattering and Harmonic Generation in Axially Symmetric Structures in COMSOL. *ACS Photoics* **2024**, *11*, 404–418.

(14) Khlebtsov, N. G. Extinction and Scattering of Light by Nonshperical Particles in Absorbing Media. *J. Quant. Spectr. Radiat. Transfer* **2022**, *280*, 108069.

(15) Olmon, R. L.; Slovick, B.; Johnson, T. W.; Shelton, D.; Oh, S.-H.; Borema, G. D.; Raschke, M. B. Optical Dielectric Function of Gold. *Phys. Rev. B* **2012**, *86*, 235147.

(16) Sönnichsen, C.; Franzl, T.; Wilk, T.; von Plessen, G.; Feldmann, J.; Wilson, O.; Mulvaney, P. Drastic Reduction of Plasmon Damping in Gold Nanorods. *Phys. Rev. Lett.* **2002**, *88*, 077402.

(17) Kreibig, U.; Genzel, L. Optical Absorption of Small Metallic Particles. *Surface Sci.* **1985**, *156*, 678–700.

(18) Novo, C.; Gomez, D.; Perez-Juste, J.; Zhang, Z.; Petrova, H.; Reismann, M.; Mulvaney, P.; Hartland, G. V. Contributions from Radiation Damping and Surface Scattering to the Linewidth of the Longitudinal Plasmon Band of Gold Nanorods: A Single Particle Study. *Phys. Chem. Chem. Phys.* **2006**, *8*, 3540–3546.

(19) Hövel, H.; Fritz, S.; Hilger, A.; Kreibig U.; Vollmer, M. Width of Cluster Plasmon Resonances: Bulk Dielectric Functions and Chemical Interface Damping. *Phys. Rev. B* **1993**, *48*, 18178-18188.

(20) Foerster, B.; Joplin, A.; Kaefer, K.; Celiksoy, S.; Link S.; Sönnichsen, C. Chemical Interface Damping Depends on Electrons Reaching the Surface. *ACS Nano* **2017**, *11*, 2886–2893.

(21) Foerster, B.; Spata, V. A.; Carte, E. A.; Sönnichsen, C.; Link, S. Plasmon Damping Depends on the Chemical Nature of the Nanoparticle Interface. *Sci. Adv.* **2019**, *5*, eaav0704.

(22) Tsai, D.-H.; DelRio, F. W.; MacCuspie, R. I.; Cho, T. J.; Zachariah, M. R.; Hackley V. A. Competitive Adsorption of Thiolated Polyethylene Glycol and Mercaptopropionic Acid on Gold Nanoparticles Measured by Physical Characterization Methods. *Langmuir* **2010**, *26*, 10325–10333.

(23) Mosquera, J.; Wang, D.; Bals, S.; Liz-Marzán, L. M. Surfactant Layers on Gold Nanorods. *Acc. Chem. Res*. **2023**, *56*, 1204−1212.

(24) Khlebtsov, N.,G.; Bogatyrev, V. A.; Khlebtsov, B. N.; Dykman, L. A.; Englebienne, P. A Multilayer Model for Gold Nanoparticle Bioconjugates: Application to Study of Gelatin and Human IgG Adsorption Using Extinction and Light Scattering Spectra and the Dynamic Light Scattering Method. *Colloid J*. **2003**, *65*, 622–635.

(25) Khlebtsov, N.G. T-matrix Method in Plasmonics: an Overview. *J. Quant. Spectrosc. Radiat.Transfer* **2013**, *123*, 184–217.

(26) Liz-Marzán, L. (Ed.). *Colloidal Synthesis of Plasmonic Nanometals*. Jenny Stanford Publishing: New York, 2021.

(27) Our definition of polarizability follows Bohren and Huffman book[7] and differs from Eq. (3) of Ref.10 by absence of host dielectric function in front of Eq. (18). This difference does not affect all conclusions as the resulted physical quantity – the induced dipole moments – are the same in both definitions.

(28) Meier, M.; Wokaun, A. Enhanced Fields on Large Metal Particles: Dynamic Depolarization. *Opt. Lett*. **1983**, *8*, 581–583

(29) Moroz, A. Depolarization Field of Spheroidal Particles. *J. Opt. Soc. Am. B* **2009**, *26*, 517–527.

(30) Jackson, J. D. *Classical Electrodynamics*, 3rd ed. Wiley: New York, 1999.

(31) Khlebtsov, N. G. Optics and Biophotonics of Nanoparticles with a Plasmon Resonance. *Quant. Electron*. **2008**, *26*, 504–529.

(32) Kreibig, U.; Volmer, M. *Optical Properties of Metal Clusters*. Springer: Berlin, 1995.

(33) Eck, D.; Helm, C. A.; Wagner, N. J.; Vaynberg, K. A. Plasmon Resonance Measurements of the Adsorption and Adsorption Kinetics of a Biopolymer onto Gold Nanocolloids. *Langmuir* **2001**, *17*, 957–960. Errata: *Langmuir* **2007**, *23*, 9522.

(34) Khlebtsov, N. G. Optical Models for Conjugates of Gold and Silver Nanoparticles with Biomacromolecules. *J. Quant. Spectrosc. Radiat.Transfer* **2004**, *89*, 143–153.

(35) Cavigli, L.; Khlebtsov, B. N.; Centi, S.; Khlebtsov, N. G.; Pini, R.; Ratto, F. Photostability of Contrast Agents for Photoacoustics: The Case of Gold Nanorods. Nanomaterials **2021**, *11*, art. 116 (1–31).

(36) Chen, H.; Shao, L.; Li, Q.; Wang J. Gold Nanorods and their Plasmonic Properties. *Chem. Soc. Rev*. **2013**, *42*, 2679–2724.

**TOC Graphic**

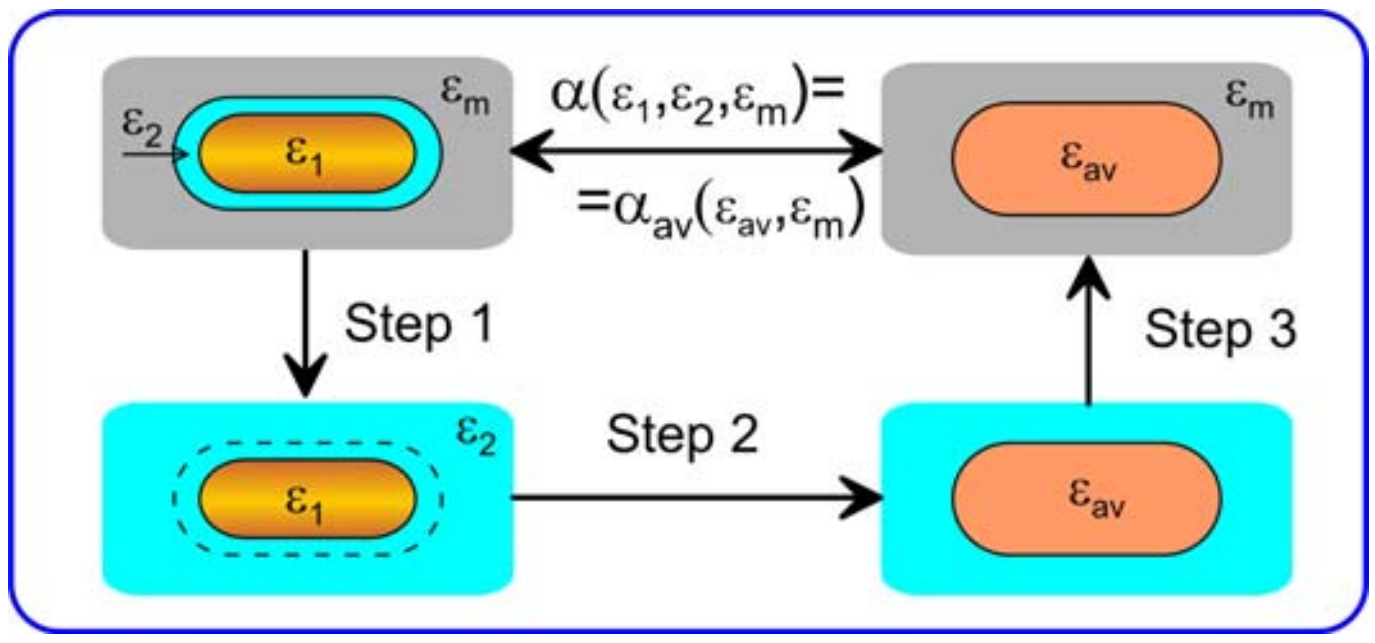

# Supporting Information

# Analytical modeling of coated plasmonic particles

Nikolai G. Khlebtsov[1,2,*], Sergey V. Zarkov[1,3]

[1]Institute of Biochemistry and Physiology of Plants and Microorganisms, "Saratov Scientific Centre of the Russian Academy of Sciences," 13 Prospekt Entuziastov, Saratov 410049, Russia

[2]SaratovStateUniversity, 83 Ulitsa Astrakhanskaya, Saratov 410012, Russia

[3]Institute of Precision Mechanics and Control, "Saratov Scientific Centre of the Russian Academy of Sciences," 24 Ulitsa Rabochaya, Saratov 410028, Russia

[*]To whom correspondence should be addressed. E-mail: (NGK) khlebtsov@ibppm.ru

**Section S1. Explicit formulas for IEA functions**

The geometrical depolarization factors $L_a$ and $L_c$ of a spheroid with semiaxes $(a, b = a, c)$ are

$$L_z = L_c = \frac{1-e_p^2}{e_p^2}\left(-1+\frac{\tanh^{-1}(e_p)}{e_p}\right),\ L_a = L_b = (1-L_c)/2, \tag{S1}$$

$e_p = \sqrt{h^2-1}/h$ is the eccentricity of a prolate spheroid with $h = c/a \geq 1$. For oblate spheroids with $h = c/a \leq 1$, the eccentricity $e_p = i\sqrt{1-h^2}/h = ie_o$ is an imaginary value, and in Eq. (S1), one has to replace $\left[\tanh^{-1}(e_p)\right]/e_p$ with $\mathrm{arctg}(e_o)/e_o$. Functions $\Omega_{1i}$ in Eq. (4) of the main text are

$$\Omega_c = \frac{9e_p^2}{25}+\frac{\varepsilon_r(1-e_p^2)-2}{5\left[1+(\varepsilon_r-1)L_c\right]}, \tag{S2}$$

$$\Omega_a = -\frac{12e_p^2}{25}+\frac{\varepsilon_r+3e_p^2-2}{5\left[1+(\varepsilon_r-1)L_a\right]}, \tag{S3}$$

where $\varepsilon_r = \varepsilon_1/\varepsilon_m$ for $F_{1i}(\varepsilon_1, \varepsilon_m)$ and $\varepsilon_r = \varepsilon_1/\varepsilon_2$ for $F_{1i}(\varepsilon_1, \varepsilon_2)$.

**Section S2. Additional Figures**

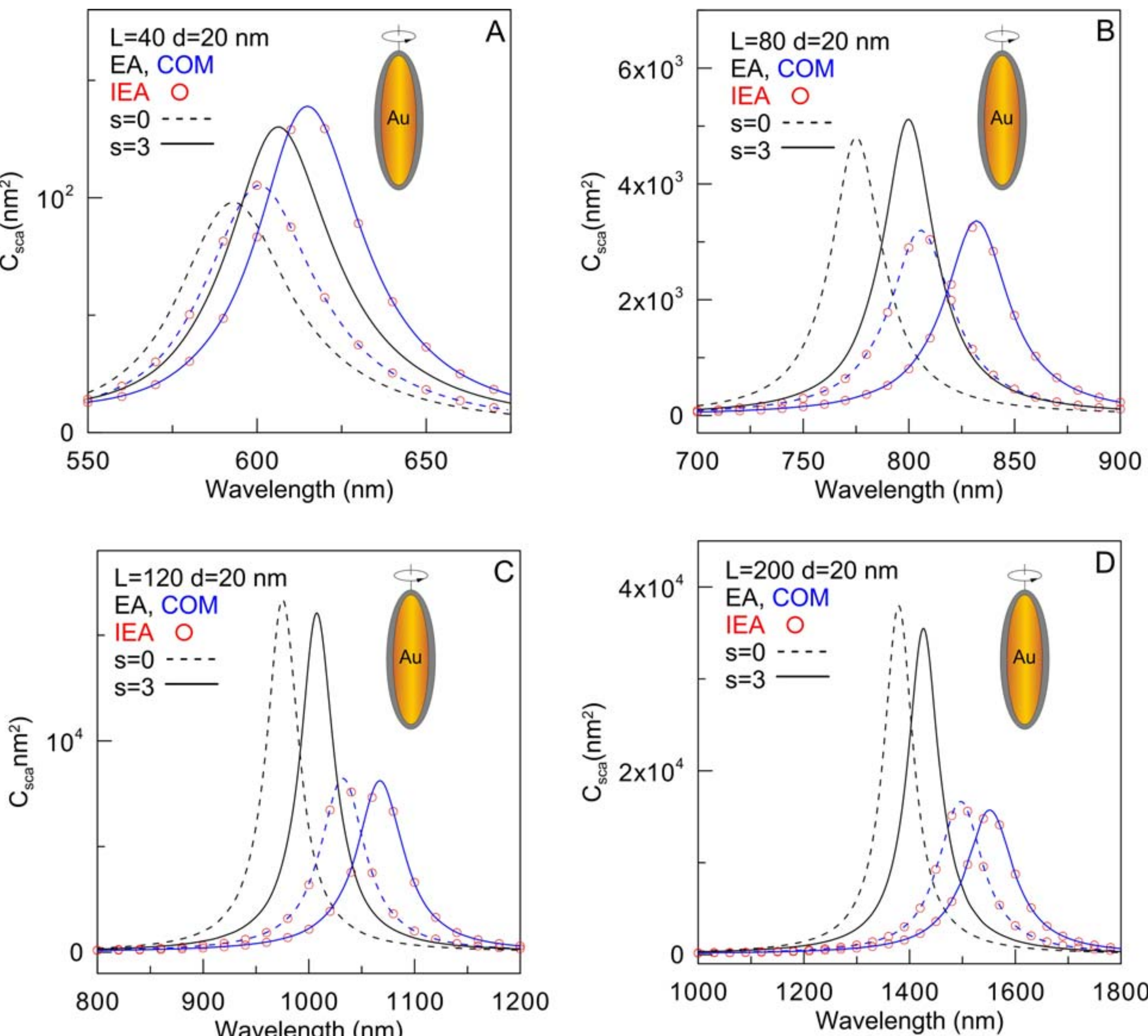

**Figure S1**. Scattering spectra for bare and coated prolate randomly oriented Au spheroids were calculated using EA (black), IEA (red circles), and COM (blue). The particle diameter is 20 nm, and the particle length is 40 (A), 80 (B), 120 (C), and 200 nm (D). The particle coating is 0 (dashed lines) and 3 nm (solid lines); the refractive index of the shell is 1.5, and the external medium is water.

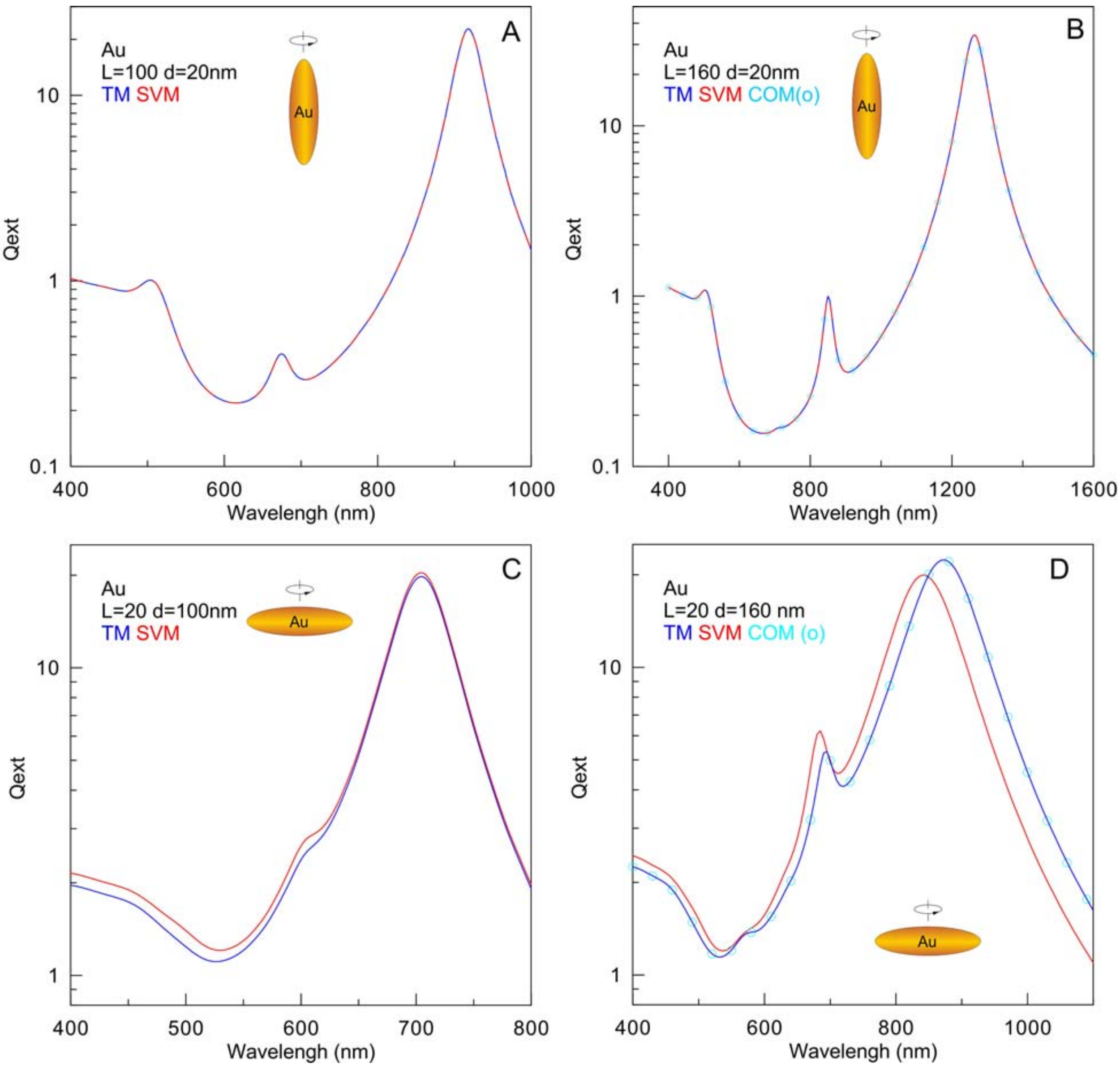

**Figure S2**. Extinction spectra of randomly oriented prolate (A and B) and oblate (C and D) Au bare spheroids with the length and diameter 100 and 20 nm (A), 160 and 20 nm (B), 20 and 100 nm (C), and 20 and 160 nm (D). Calculations by TM (blue), SVM (red), and COMSOL (cyan circles). It is evident that the SVM data for oblate spheroids are inaccurate whereas TM and COM spectra coincide.

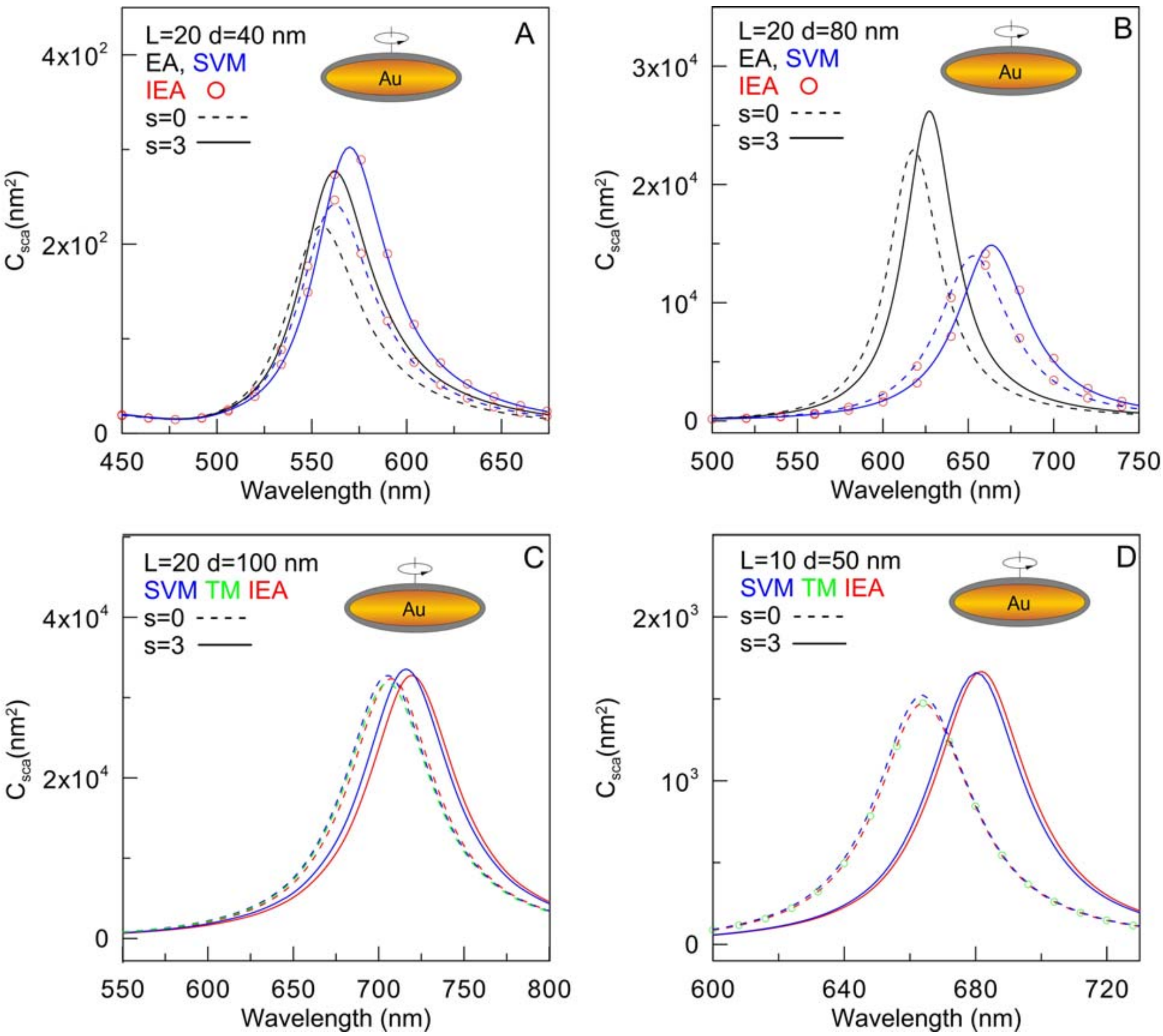

**Figure S3**. Scattering spectra for bare and coated oblate randomly oriented Au spheroids calculated by EA (black), SVM (blue), and IEA (red). The particle length $L = 2c_1$ is 20 nm (A-C) and 10 nm (D). The particle diameter $d = 2a_1$ is 40 (A), 80 (B),100 (C), and 50 nm (D). The particle coating thickness is 0 (dashed lines) and 3 nm (solid lines); the refractive index of the shell is 1.5, and the external medium is water. Note that for small bare ($s = 0$) particles with $L = 2c_1$=10 nm and $d = 2a_1 = 50$ nm, SVM and TM give identical spectra (D, red dashed line, and green circles); however, for larger 20x100 spheroids with the same aspect ratio $a / c = 5$, SVM data are somewhat inaccurate (panel C, green (TM simulations) and blue dashed lines).

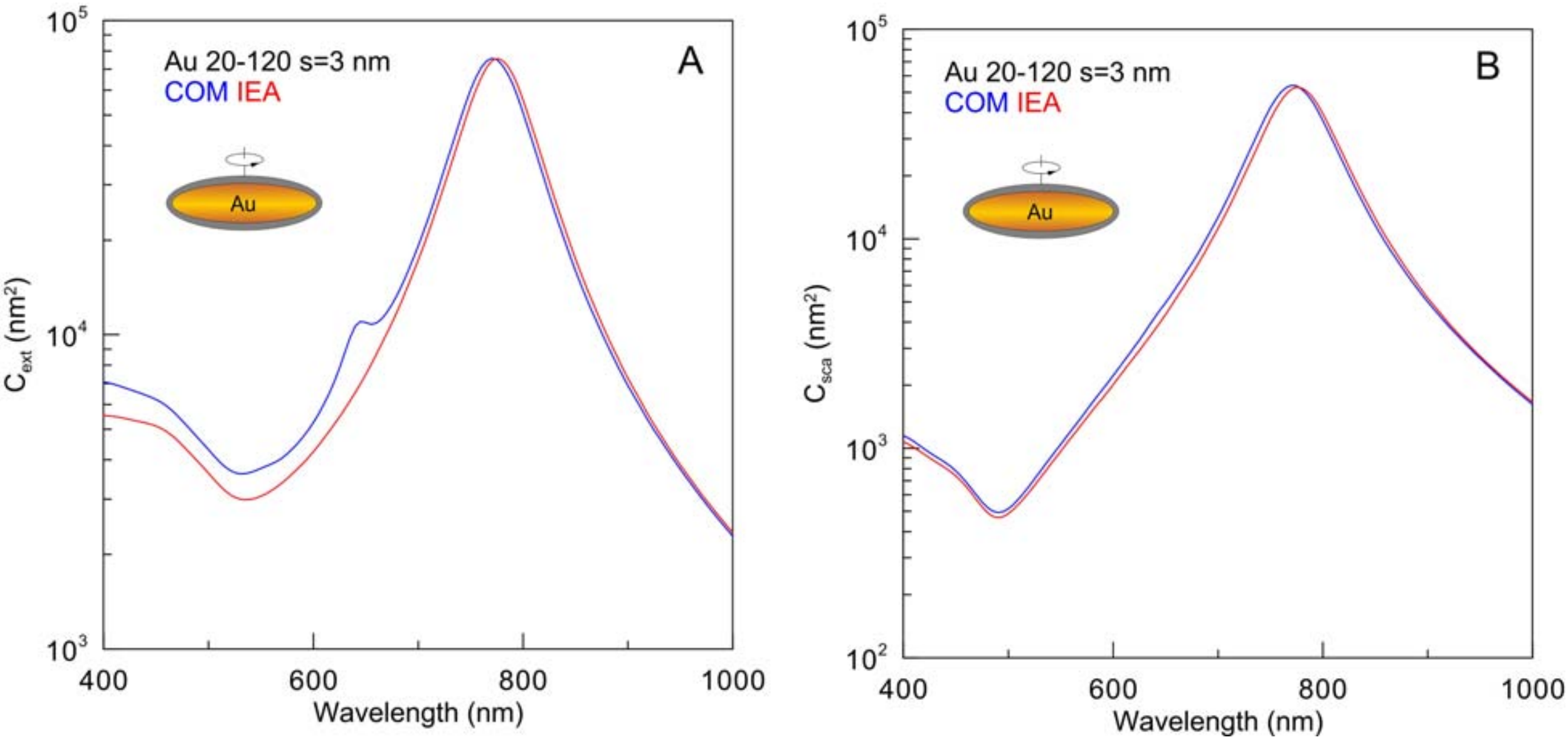

**Figure S4**. Extinction (A) and scattering (B) spectra for bare and coated oblate randomly oriented Au spheroids were calculated using COMSOL (blue) and IEA (red). The particle length $L = 2c_1$ and diameter are 20 and 120 nm, respectively. The particle coating thickness is 3nm, the refractive index of the shell is 1.5, and the external medium is water. Note the absence of the quadrupole peak in panel B. Thus, this qudarupole peak in the extinction spectrum (panel A) is related to the absorption only.

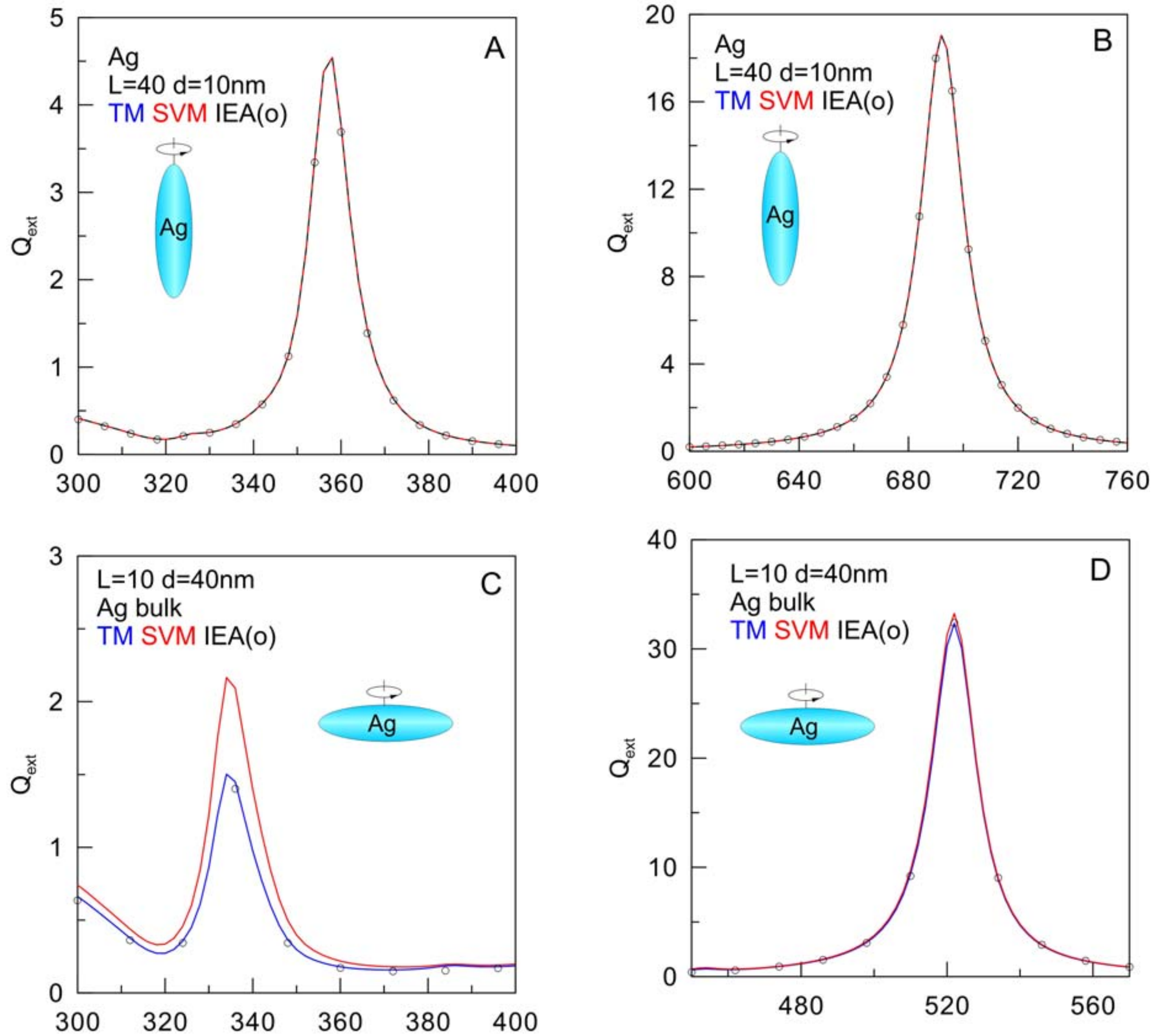

**Figure S5**. Extinction spectra for bare prolate (A, B) and oblate (C, D) randomly oriented Ag spheroids calculated by TM (black), SVM (red), and IEA (black). The particle length $L = 2c_1$ and diameter are 40 and 10 nm (A, B) and 10 and 40 nm (C, D), respectively. Note that SVM data for oblate particles are inaccurate for wavelengths less than 400 nm, whereas IEA data agree well with TM.

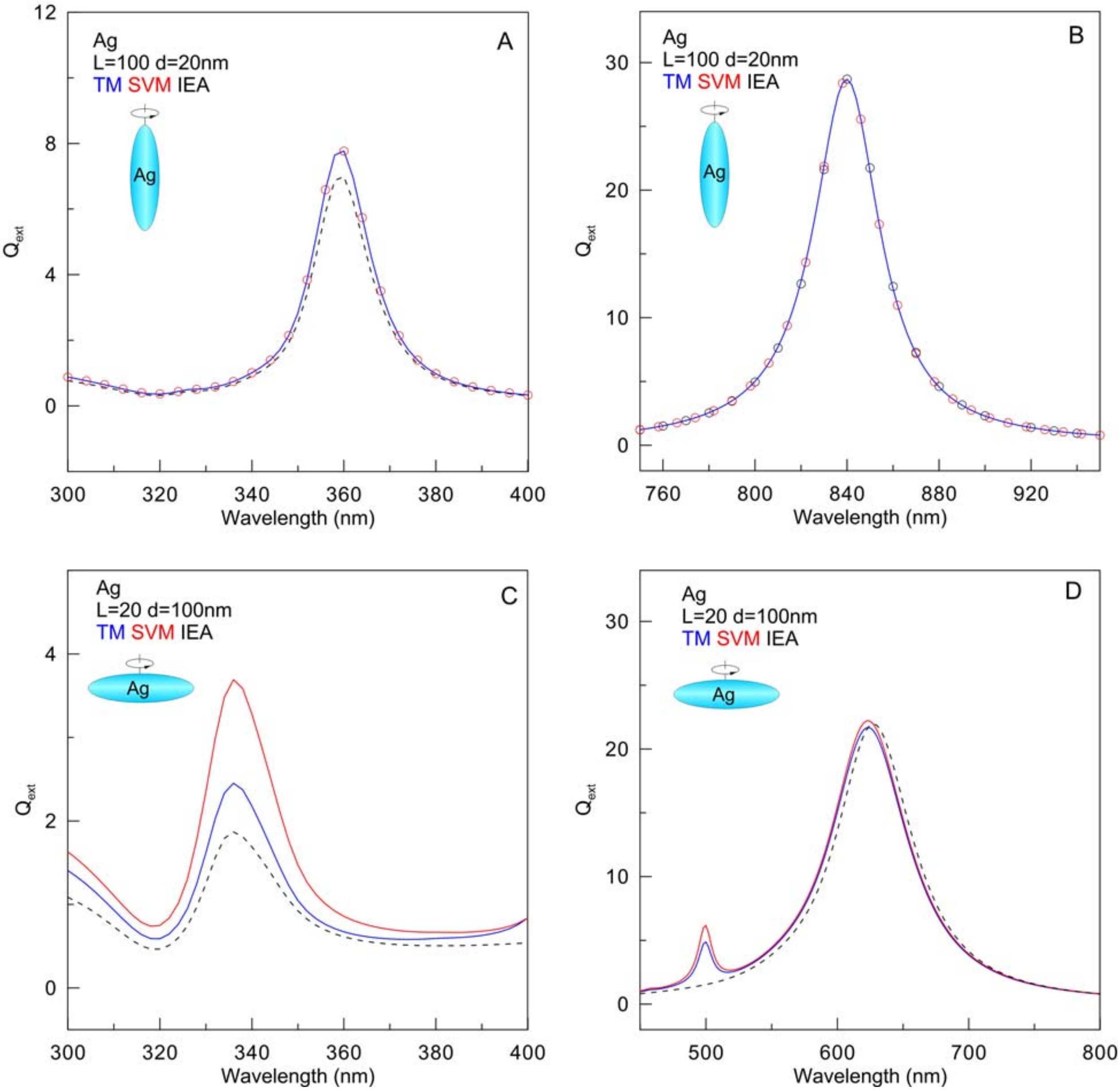

**Figure S6**. Extinction spectra for bare prolate (A, B) and oblate (C, D) randomly oriented Ag spheroids calculated by TM (black), SVM (red), and IEA (black). The particle length $L = 2c_1$ and diameter $d = 2a_1$ are 100 and 20 nm (A, B) and 20 and 100 nm (C, D), respectively. Note that for oblate particles, both SVM and IEA data are inaccurate for wavelengths less than 400 nm.

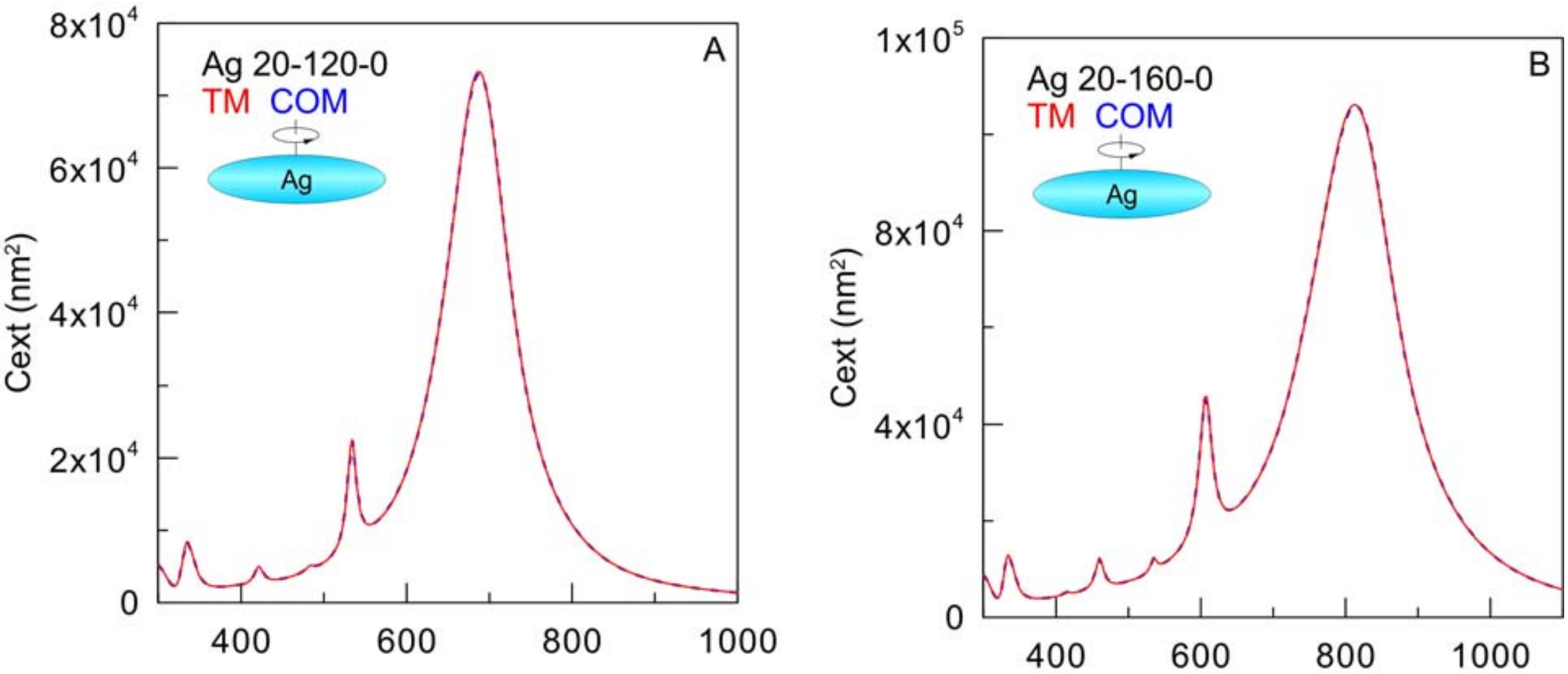

**Figure S7**. Extinction spectra for oblate randomly oriented Ag spheroids. Calculations by TM (red) and COM (blue). The particle diameter $d = 2a_1$ is 20 nm, and the particle length $L = 2c_1$ is 120 (A) and 160 nm (B). Both methods give identical plots.

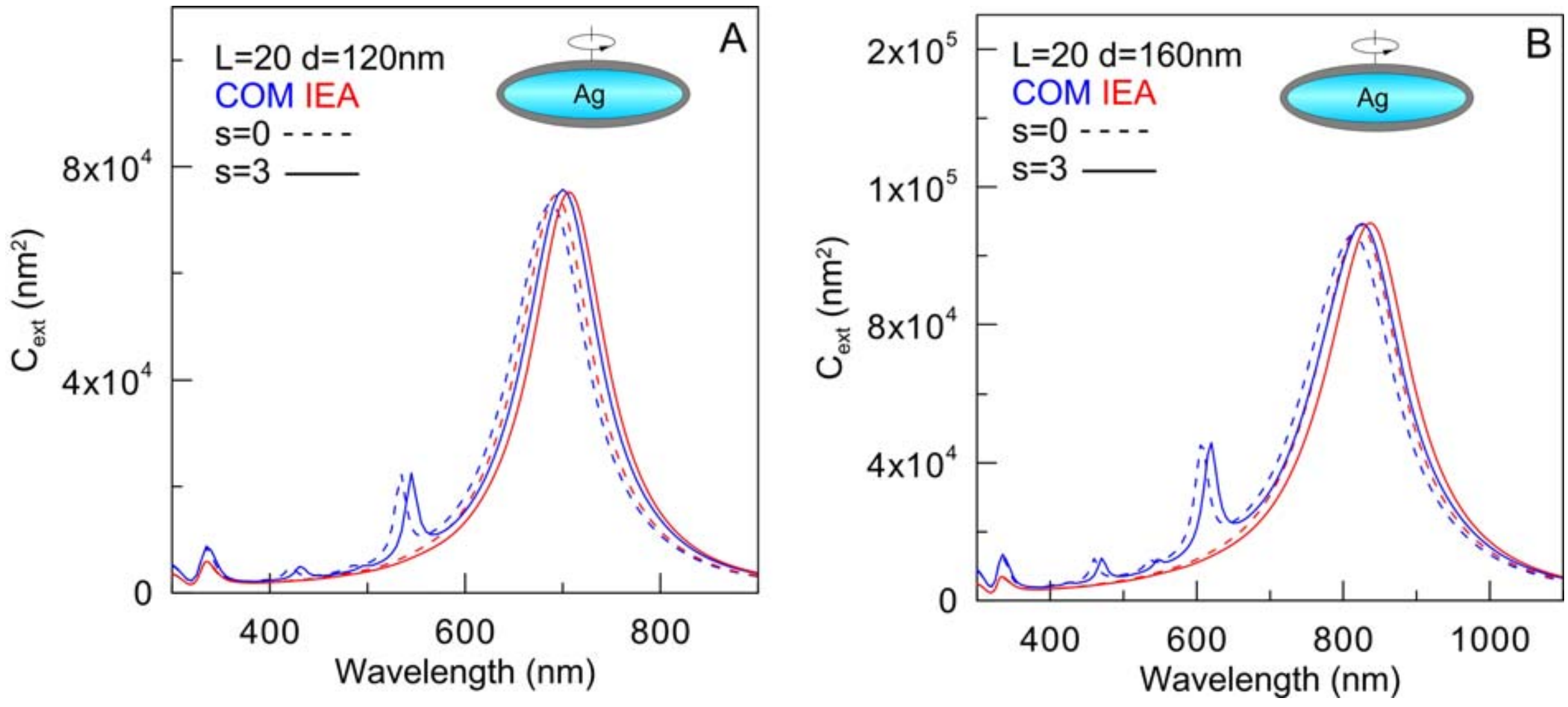

**Figure S8**. Extinction spectra of randomly oriented Ag bare and coated randomly oriented spheroids were calculated using COMSOL (blue) and IEA (red). The particle diameter is 20 nm, and their length is 120 (A) and 160 nm (B). The coating thickness is 0 (dashed lines) and 3 nm (solid line); the refractive index of the coating is 1.5.

## Section S3. Optical constants of bulk silver and water

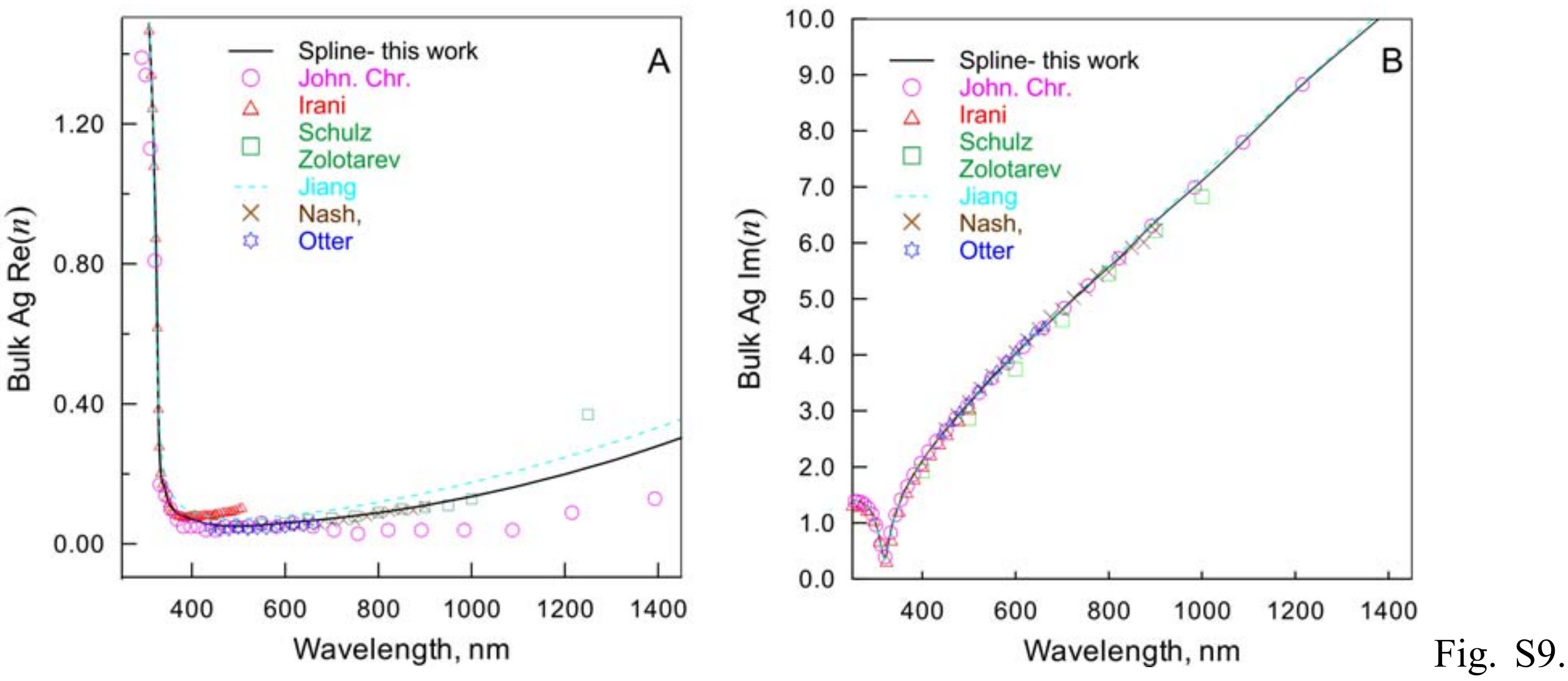

**Figure S9**. Spectral dependence of the bulk Ag refractive index taken from several literature data[1-7] and a cubic spline approximation (black solid line).

The refractive index of water was calculated by the equation

$$n_m = 1.3233 + \frac{3.478\times10^{-3}}{\lambda^2} - \frac{5.111\times10^{-5}}{\lambda^4},\ [\lambda] = \mu\text{m}\ . \qquad \text{(S4)}$$

**Section S4. Gallery of plasmonic nanoparticles fabricated at IBPPM RAS.**

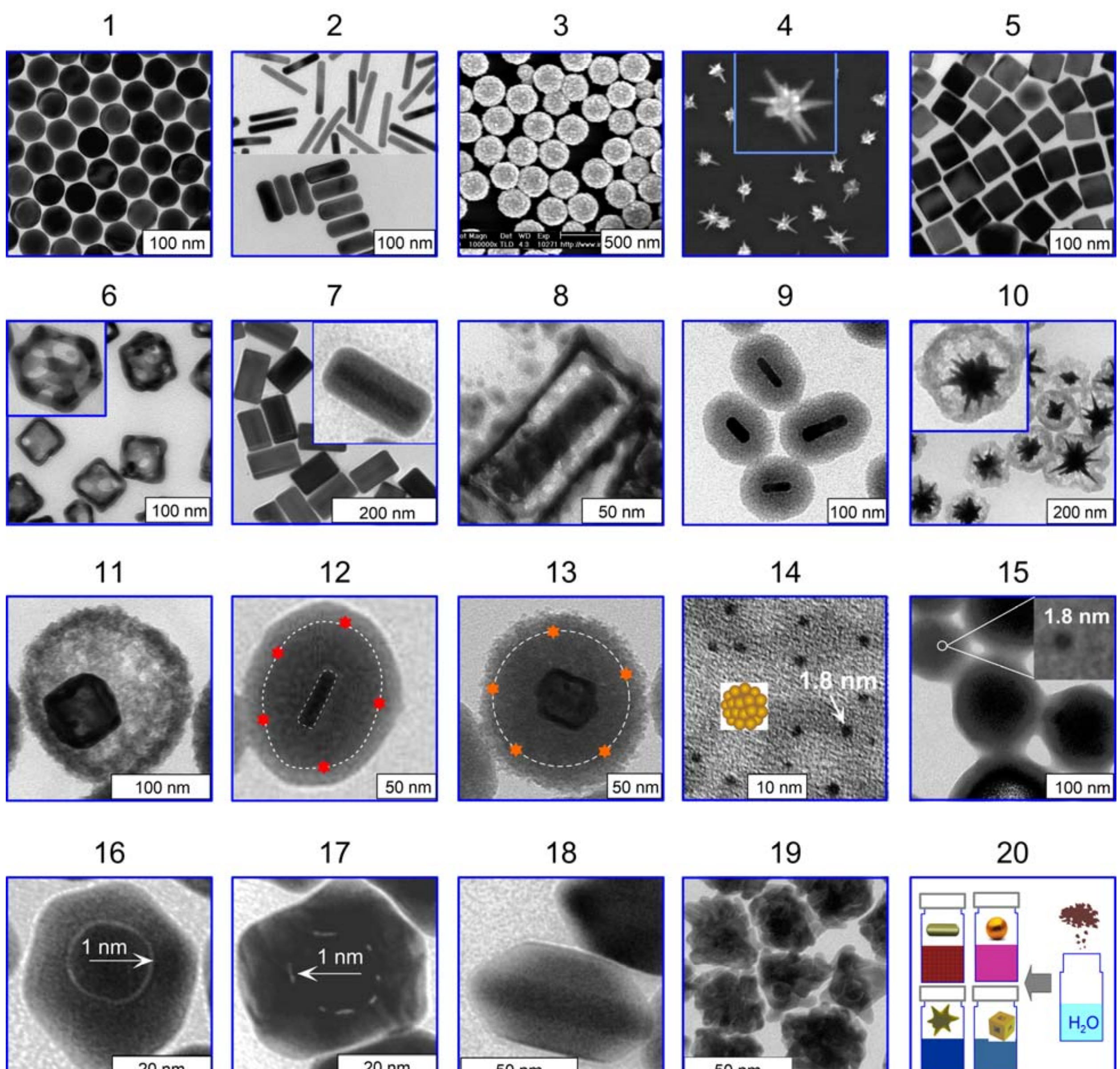

**Figure S10**. TEM gallery of images of nanoparticles and nanocomposites synthesized at the Laboratory of Nanobiotechnology, IBPPM RAS: gold nanospheres (1); gold nanorods (2); gold nanoshells on silica cores (3); gold nanostars (4 ); silver nanocubes (5) used as templates for the synthesis of gold nanocages ( 6 ); nanocuboids composed of gold nanorods coated by a silver shell (7); anisotropic gold nanocages obtained from nanocuboids (8); gold nanorods (9 ), nanostars ( 10 ) and nanocages (11) coated by a mesoporous silica shell; gold nanorods (l2 ) and nanocages (13) coated by a conventional and mesoporous silica shell and doped with photodynamic agents; fluorescent atomic gold nanoclusters Au25 stabilized by bovine serum albumin (BSA) molecules (14); human serum albumin (HSA) nanoparticles doped with fluorescent atomic clusters (15); SERS gold labels with benzene-1,4-dithiol molecules embedded into a nanometre hollow (16) or bridged (17) gap between the spherical (16) or polygonal (17)

core and the shell; anisotropic labels with a gold nanorod as the core and benzene-1,4-dithiol or nitrobenzene-1,4-dithiol embedded into the 1-nm gap between the core and the gold shell (18); SERS labels based on gold nanorods functionalized by 4-aminothiophenol and 4-nitrobenzenethiol molecules coated by a silver shell (19); water-soluble powders of gold nanospheres, nanorods, nanostars and nanocages (20) (powder particles coated by thiolated mPEG-SH molecules, Mw=5000). For related references, see Ref.[8].

## References

(1) Schulz, L. G. The Optical Constants of Silver, Gold, Copper, and Aluminum. I. The Absorption Coefficient. *J. Opt. Soc. Am.* **1954**, *44*(5), 357–361.

(2) Schulz, L. G.; Tangherlini, R. F. The Optical Constants of Silver, Gold, Copper, and Aluminum. II. The Refraction coefficient. *J. Opt. Soc. Am.* **1954**, *44*(5) 362–368.

(3) Otter M. Optische Konstanten massiver Metalle. Z. Physik. 1961, 161(1-2), 163–178.

(4) Jiang, Y.; Pillai, S.; Green, M. A. Re-Evaluation of Literature Values of Silver Optical Constants. *Opt. Express* **2015**, *23*(3), 2133–2144.

(5) Nash, D. J.; Sambles, J. R. Surface Plasmon-Polariton Study of the Optical Dielectric Function of Silver. *J. Mod. Opt.* **1996**, *43*(1), 81–91.

(6) Irani, G. B.; Huen, T.; Wooten F. Optical Constants of Silver and Gold in the Visible and Vacuum Ultraviolet. *J. Opt. Soc. Am.* **1971**, *61*, 128–129.

(7) Johnson, P. B.; Christy, R. W. Optical Constants of the Noble Metals. *Phys. Rev. B* **1972**, 6, 4370–4379.

(8) Dykman, L.A.; Khlebtsov, N. G. Methods for chemical synthesis of colloidal gold. *Russ. Chem. Rev.* **2019**, *88*(3), 229–247.